\pdfoutput=1
\documentclass[journal=jacsat,manuscript=article]{achemso}

\usepackage[version=3]{mhchem} 

\usepackage{graphicx}

\usepackage{color}
\usepackage{dcolumn} 
\usepackage{bm}      
\usepackage{amssymb}
\usepackage{amsfonts}
\usepackage{svg}
\usepackage{comment}
\usepackage{booktabs} 
\usepackage{xr}
\usepackage[bookmarks=false,colorlinks,citecolor=red]{hyperref}
\usepackage{float}
\usepackage{caption}
\usepackage{amsmath}
\usepackage{mhchem}
\usepackage{subcaption}
\usepackage[colorlinks,citecolor=red]{hyperref}
\usepackage{multicol}
\usepackage{multirow}
\usepackage{algorithm}
\usepackage{algpseudocode}
\usepackage{hyperref}
\usepackage[normalem]{ulem}
\usepackage[symbol]{footmisc}
\SectionNumbersOn

\newcommand{\aidag}[1]{\hat{a}_{#1}^{\dag}}
\newcommand{\ai}[1]{\hat{a}_{#1}}

\newcommand{\COMMENTED}[1]{}

\makeatletter
\newcommand*{\addFileDependency}[1]{
  \typeout{(#1)}
  \@addtofilelist{#1}
  \IfFileExists{#1}{}{\typeout{No file #1.}}
}
\makeatother

\newcommand*{\myexternaldocument}[1]{%
    \externaldocument{#1}%
    \addFileDependency{#1.tex}%
    \addFileDependency{#1.aux}%
}

\myexternaldocument{supporting-no-color}

\author{Prateek Vaish}
\affiliation{Department of Chemistry, Brown University, Providence, Rhode Island 02912, USA}
\author{Brenda M. Rubenstein}
\affiliation{Department of Chemistry, Brown University, Providence, Rhode Island 02912, USA}
\affiliation{Department of Physics, Brown University, Providence, Rhode Island 02912, USA}
\affiliation{Data Science Institute, Brown University, Providence, Rhode Island 02912, USA}
\email{brenda_rubenstein@brown.edu}

\title[]
  {Reducing the Cost of Unitary Coupled Cluster via Active Space Partitioning}

\begin{document}

\begin{abstract}
Unitary Coupled Cluster (UCC) theory is a promising variational method for electronic structure calculations, particularly for systems that exhibit strong electronic correlation and for implementation on quantum computers. However, its practical application is limited to small chemical systems with small basis sets due to its steep computational scaling that results from its non-terminating Baker--Campbell--Hausdorff expansion. Here, we introduce an active space UCCSD(4)/MP2 approach that leverages a fourth-order many-body perturbation theory truncation of UCCSD within a selected active space, while treating external excitations at the MP2 level. We explore two variants: a composite method that sums separate internal and external contributions and an interacting method that couples the amplitudes for potentially greater accuracy. We test our approach on a range of systems, including molecules from the GW100 dataset in their equilibrium geometries, a moderately correlated metaphosphate hydrolysis reaction, and the strongly correlated torsion of ethylene. Our results suggest that the interacting method with canonical orbitals is robust and stable for both weakly and moderately correlated systems and accurately reproduces the full UCCSD(4) potential energy curves including only 15--25\% of the virtual orbitals in its active space. In comparison, the composite formulation exhibits greater sensitivity to the choice of orbital basis and active space size, leading to less systematic behavior across the benchmark set. For ethylene torsion, a system dominated by strong static correlation, both composite and interacting formulations employing canonical orbitals closely track the full UCCSD(4) reference but do not alleviate the unphysical features inherited from the underlying single-reference UCCSD(4) description. This active space framework offers a computationally tractable approach for modeling correlated molecules and reactions on classical computers and provides a viable path for scaling UCC calculations for resource-constrained quantum hardware.
\end{abstract}


\section{Introduction}
Coupled cluster (CC) theory has emerged as one of the most accurate theories for describing the electronic structure of molecules \cite{bartlett2007coupled,bartlett2024perspective,bartlett2012coupled} and materials \cite{mcclain2017gaussian,zhang2019coupled}. Through its relatively compact exponential ansatz of singles, doubles, triples, and even higher-order electron excitations, coupled cluster theory can describe important electron correlations with fewer explicit terms than other comparable wave function theories. As a result, coupled cluster has become the ``gold standard" electronic structure method in quantum chemistry that has been successfully leveraged to understand a range of chemical reactions~\cite{sparta2014chemical}, nanoclusters~\cite{shi2024going}, and most recently, correlated solids~\cite{gruber2018applying,neufeld2023highly}.

Nonetheless, traditional coupled cluster theories exhibit steep scaling (e.g., $O(N^6)$ for CCSD and $O(N^8)$ for CCSDT) with system size and face challenges accurately describing bond breaking and other multireference problems~\cite{helgaker2013molecular,bartlett2007coupled}. The core of this limitation lies in the single reference nature of the CC ansatz, which builds upon a single Slater determinant (usually Hartree Fock wavefunction) as its starting point. This approach is highly effective when a single Slater determinant provides a good zero-order description. However, in multireference problems--such as stretching a chemical bond to dissociation~\cite{sherrill_bond_2007}, describing diradicals~\cite{bertels_accurate_2014}, or handling certain excited states~\cite{dreuw_single-reference_2005}--the electronic structure becomes poorly described by any single configuration. Instead, a linear combination of two or more determinants is essential for even a qualitatively correct picture. In such cases, the standard CC methods, which are based on a qualitatively incorrect reference, struggle to recover the proper physics, leading to unphysical potential energy surfaces and incorrect dissociation energies. Moreover, traditional coupled cluster theory employs non-unitary operators that cannot straightforwardly be implemented on emerging quantum architectures that require unitary operators.\cite{motta2022emerging}

These challenges invite the development of alternative versions of coupled cluster theory. One promising theory along these lines is Unitary Coupled Cluster (UCC) theory, a version of coupled cluster theory that is entirely constructed from unitary operators.\cite{taube2006new} Unlike traditional coupled cluster methods, UCC is variational. As a result, it avoids the non-variational collapse of the energy that occurs in traditional, non-variational coupled cluster theory when stretching bonds~\cite{mizukami_orbital_2020}. In UCC, the wavefunction is parameterized as $e^{\hat{T} - \hat{T}^\dagger} |\Psi_0\rangle$, where $\hat{T}$ is the cluster operator, and the amplitudes are optimized variationally, often using gradient-based methods on classical computers or hybrid quantum-classical approaches. Because of its unitary nature, UCC has emerged as the leading electronic structure method employed on quantum computers within the Variational Quantum Eigensolver framework or for initializing wave functions for other quantum algorithms~\cite{anand2022quantum,evangelista2019exact,lee_generalized_2019,romero2018strategies}. Recently, UCC has been implemented on quantum hardware to study small molecules like H$_2$, LiH, and BeH$_2$, demonstrating its potential for accurate energy calculations despite noise~\cite{guo2024experimental}. Nonetheless, the unitary nature of UCC also results in an operator expansion that, unlike that in conventional coupled cluster theory, does not naturally truncate. UCC must thus typically be truncated at a given order, and even with truncation, can be far more expensive than CC~\cite{evangelista_alternative_2011}. Accordingly, practical implementations of UCC on quantum computers to model realistic chemical systems typically necessitate the use of more operators, and correspondingly gates, than are or will be available in the foreseeable future.

In recent years, a number of methods have been designed to reduce the practical cost of UCC theory. On classical computers, strategies have largely focused on creating tractable approximations. A primary approach involves truncating the non-terminating Baker-Campbell-Hausdorff expansion, as explored in early work by Bartlett and coworkers~\cite{Bartlett_Kucharski_Noga_1989, kutzelnigg_error_1991}. More recent efforts have focused on approaches like stochastic unitary coupled cluster theory, which enables efficient sampling of excitation amplitudes~\cite{filip_stochastic_2020}; perturbative triples corrections like UCCSD(T), which augment truncated UCCSD with non-perturbative triples for improved accuracy~\cite{Windom_Claudino_Bartlett_2024}; and a comparison study by DePrince and coworkers that evaluated perturbative versus commutator-rank-based truncation schemes for UCCSD highlighting trade-offs in accuracy and efficiency~\cite{phillips_comparing_nodate}. On the quantum side, adaptive approaches like the Adaptive Derivative-Assembled Pseudo-Trotter
ansatz-Variational Quantum Eigensolver (ADAPT-VQE),\cite{grimsley2019adaptive,tang2021qubit} iteratively selects important operators based upon energy gradients; generalized methods like k-UpCCGSD produce shallower circuits~\cite{lee_generalized_2019}; and orbital optimization techniques to enhance convergence~\cite{mizukami_orbital_2020, sokolov_quantum_2020}. In tandem with these developments, embedding and active space partitioning methods have gained prominence for their ability to focus computational resources on strongly correlated regions of large systems. Techniques such as Local Active Space Unitary Coupled Cluster (LAS-UCC)~\cite{otten_localized_2022}, Density Matrix Embedding Theory (DMET)~\cite{knizia_density_2012}, and Bootstrap Embedding~\cite{ye_bootstrap_2019} provide a framework for capturing essential physics by treating localized fragments at a high level of theory while embedding them in a more efficiently described environment. The choice of a one-electron basis is also critical as it influences the efficiency of the UCC expansion. Canonical orbitals (COs), the eigenfunctions of the Fock operator, are the standard choice for orbitals, but they provide a delocalized representation of electron correlation, potentially slowing the convergence of the UCC expansion. Frozen Natural Orbitals (FNOs), which are obtained by diagonalizing the one-particle reduced density matrix from a correlated calculation (typically MP2) and truncating the virtual space based on occupation numbers, offer a compelling alternative~\cite{sosa_selection_1989}. This approach exploits the fact that a significant portion of dynamic correlation is captured by a relatively small subset of virtual orbitals. By retaining only the most occupied virtual orbitals, FNOs provide a more compact representation of the wavefunction than COs, potentially capturing essential physics with a significantly reduced orbital space. For UCC on quantum computers, this is particularly useful, as a compact orbital space can reduce the number of qubits as well as the circuit depth. While COs ensure a diagonal Fock operator, a desirable property for perturbation theory, the study by Lange and Berkelbach~\cite{Lange_Berkelbach_2020} has demonstrated that natural orbitals often yield superior accuracy for active space methods. Specifically, they found that in their CCSD-based active space framework, NOs frequently provided better agreement with high-level benchmarks like CCSD(T) compared to COs, as the compact NO basis recovers a larger fraction of the correlation energy within the truncated space.

In this manuscript, we introduce an active space Unitary Coupled Cluster approach, UCCSD(4)/MP2, designed to model electron correlation in chemical systems with significantly reduced computational overhead. By leveraging a classical UCCSD(4) approach~\cite{Bartlett_Kucharski_Noga_1989}, derived from a fourth-order many-body perturbation theory (MBPT) truncation of the UCC energy functional, we establish a balanced framework that preserves the extensivity and variational nature of the unitary formulation while avoiding the cost of a non-terminating Baker-Campbell-Hausdorff expansion. This hybrid method partitions the cluster operator into internal and external parts, treating critical correlations within a defined active space using UCCSD(4) and the remaining external space via efficient, non-iterative MP2 calculations. We demonstrate the performance of this approach on a variety of small molecules in their equilibrium geometries, bond-breaking reactions, including phosphate hydrolysis, and the torsion of ethylene, which have varying degrees of electron correlation. We moreover investigate the impact of orbital choice and the method for coupling the active and external spaces on accuracy and stability. Notably, our findings indicate that COs are consistently more efficient and reliable than MP2--based NOs for our active space approach. We attribute this to the fact that canonical orbitals preserve the diagonal zeroth-order Hamiltonian required by the underlying many-body perturbation theory partitioning, whereas the off-diagonal Fock matrix elements inherent to the natural orbital basis violate the assumptions of the UCCSD(4) truncation, leading to numerical instability. Our results illustrate that our approach may offer an accurate framework for modeling strongly correlated molecules and reactions at a reduced cost, which is of value for the computationally efficient modeling of correlated reactions on classical computers. Furthermore, this framework can be adapted to scale quantum UCC calculations to larger molecules and basis sets, enhancing its applicability on resource-limited quantum hardware. 

\section{Methods}
\label{sec:methods}
In this section, we describe the theoretical framework underlying our active space unitary coupled cluster approach. We begin with a brief overview of traditional coupled cluster theory and its unitary variant, followed by truncation schemes based on many-body perturbation theory. We then introduce the active space approximation with both canonical and natural orbitals and its integration with MP2 for efficient computations.

\subsection{Theoretical Framework}

\subsubsection{Unitary Coupled Cluster}
The coupled cluster wavefunction, $|\Psi_{CC}\rangle$, is obtained by applying an exponential of a cluster operator onto a reference wavefunction, $|\Psi_0\rangle$, usually the Hartree-Fock wavefunction~\cite{bartlett2007coupled,helgaker2013molecular} 
\begin{equation}
   \vert \Psi_{CC} \rangle = e^{\hat{T}} \vert  \Psi_0 \rangle,
\end{equation}
where the cluster operator, $\hat{T}$, can be written as a sum of excitation operators
\begin{equation}
    \label{"eq:cluster operator"}
    \hat{T} = \sum_{k = 1}^{N} \hat{T_k}.
\end{equation}
$\hat{T_k}$ represents the degree of the excitation operator applied to the reference wavefunction. In principle, the full CC algorithm is exact and has a cost that scales exponentially with system size. Therefore, in practice, the cluster operator is truncated considering only the lowest few excitations depending on the system size and available computational resources. For CCSD, $\hat{T} = \hat{T_1} + \hat{T_2}$, where the single excitation operator is given by $\hat{T_1} = \sum\limits_{a i} t_i^a \aidag{a} \ai{i}$ and the double excitation operator is given by $\hat{T_2} = \frac{1}{4} \sum\limits_{ijab} t_{ij}^{ab} \aidag{a} \aidag{b} \ai{j} \ai{i}$. $t_i^a \text{ and } t_{ij}^{ab}$ are the unknown cluster amplitudes that need to be computed to obtain the coupled cluster energy. The cluster amplitudes, \( t_i^a \) and \( t_{ij}^{ab} \), are determined by projecting the Schrödinger equation,

\begin{equation}
    e^{-\hat{T}} \hat{H} e^{\hat{T}} \vert \Psi_0 \rangle = E_{CC} \vert \Psi_0 \rangle,
\end{equation}
onto the excited determinants \( \langle \Phi_i^a \vert \) and \( \langle \Phi_{ij}^{ab} \vert \), leading to a system of nonlinear equations that are solved iteratively~\cite{purvis_full_1982}.

In the traditional coupled cluster algorithm, the Hamiltonian is similarity transformed as $\bar{H} = e^{-\hat{T}} \hat{H} e^{\hat{T}}$, which has a naturally truncating Baker-Campbell-Hausdorff (BCH) expansion. The transformed Hamiltonian is no longer Hermitian; therefore, the expectation value of the energy does not obey the variational theorem. Loss of variationality is not a concern in most cases; however, in systems in which a single reference wavefunction like the Hartree Fock determinant is a poor approximation, the energy may diverge towards negative infinity in strongly correlated regimes~\cite{bochevarov_hybrid_2005}. 

The Unitary Coupled Cluster method addresses this problem by using a wavefunction of the form
\begin{equation}
\mid \Psi_{UCC} \rangle=  e^{\hat{\tau}}  \mid \Psi_0\rangle, \quad \text { for }\langle\Psi_{UCC} \mid \Psi_0 \rangle \neq 0 \text {, }
\end{equation}
where $\hat{\tau} = \hat{T} - \hat{T^\dagger}$ is an anti-Hermitian operator.  $\hat{T}$ is the cluster (excitation) operator defined in Equation \ref{"eq:cluster operator"} and $\hat{T^\dagger}$ is the de-excitation operator. As a result, the exponential operator $e^{\hat{\tau}}$ is unitary. The energy expectation value,
\begin{equation}
    E_{UCC}=\frac{\left\langle \Psi_0\left|e^{\hat{\tau}^\dagger} \hat{H} e^{\hat{\tau}}\right| \Psi_0\right\rangle}{\left\langle \Psi_0\left|e^{\hat{\tau}^\dagger} e^{\hat{\tau}}\right| \Psi_0\right\rangle} = \left\langle \Psi_0\left|e^{\hat{\tau}^\dagger} \hat{H} e^{\hat{\tau}}\right| \Psi_0\right\rangle
\end{equation}
is symmetric, and therefore, obeys the variational theorem. Similar to traditional CC, the energy in UCC can be written in connected cluster form,
\begin{equation}
    \label{eq:ucc energy connected cluster}
    E_{UCC} = \langle \Psi_0 \mid (\hat{H} e^{\hat{\tau}})_{C} \mid \Psi_0 \rangle,
\end{equation}
making it extensive. 

The UCC method overcomes the challenges faced by traditional CC by preserving the variational property while remaining extensive. However, the BCH expansion of the transformed Hamiltonian in UCC no longer terminates naturally:
\begin{equation}
    \label{eq:UCC BCH expansion}
    e^{-\hat{\tau}} \hat{H} e^{\hat{\tau}} \approx  \  \hat{H} + [\hat{H}, \hat{\tau}] +  \frac{1}{2}[[\hat{H}, \hat{\tau}], \tau] + \frac{1}{6}[[[\hat{H}, \hat{\tau}], \tau], \hat{\tau}] + ...
\end{equation}
This is due to the presence of mixed terms like $\left[\hat{H},\left[T, T^{\dagger}\right]\right]$ in the expansion. Due to the non-truncating expansion of the energy (and amplitude) equations, the method scales exponentially with system size. Several truncation schemes have been proposed in the literature to circumvent this problem~\cite{Bartlett_Kucharski_Noga_1989, kutzelnigg_error_1991, taube_new_2006, evangelista_alternative_2011, liu_quadratic_2022}. The most straightforward approach is the truncation of the exponential expansion of the wavefunction to an arbitrary order~\cite{filip_stochastic_2020, evangelista_alternative_2011}. When substituting this into the UCC energy functional given by Equation~\ref{eq:ucc energy connected cluster}, one can stop at a finite order of expansion to achieve a practical balance between computational cost and accuracy. Typically, convergence in the energy to the third decimal place can be attained by truncating the series to the fourth or fifth order. This approach can be implemented in a Configuration Interaction (CI) or determinant-based code, though mixed terms such as $T^{\dagger} H T$ still lead to poor scaling due to a large number of non-canceling Wick contractions. Another approach is to truncate the Baker-Campbell-Hausdorff (BCH) expansion in Equation~\ref{eq:UCC BCH expansion} by cutting off at a certain truncation order. Similar to exponential-based truncation, this BCH-based method can be implemented in CI or determinant-based frameworks to obtain a tractable approximation at the expense of some accuracy~\cite{liu_quadratic_2022}. In both these cases, the quality of the reference determinant plays a crucial role: if the chosen reference is good and the cluster amplitudes are small, then higher-order terms contribute less significantly and truncation schemes become more reliable.

\subsubsection{Truncation via Many-Body Perturbation Theory}

A useful perspective for designing truncation schemes comes from many-body perturbation theory (MBPT)~\cite{Bartlett_Kucharski_Noga_1989}. In this approach, the normal ordered Hamiltonian $H_N$ is partitioned into a zeroth-order Fock operator $f_N$ and a first-order perturbation operator $W_N$ containing two-body interactions: 
\begin{equation}
    \hat{H}_N = \hat{f}_N + \hat{W}_N,
\end{equation}
where $\hat{f}_N = \sum_p \epsilon_p {\hat{p}^\dagger \hat{p}}$ and $\hat{W}_N = \frac{1}{4} \sum_{pqrs} \langle pq || rs \rangle {\hat{p}^\dagger \hat{q}^\dagger \hat{s} \hat{r}}$. Within this framework, the UCC equations can be truncated to a given order in $\hat{W}_N$. By working to fourth order in the energy functional and systematically including terms involving $T_1$ and $T_2$ amplitudes only, one obtains a set of tractable equations for the amplitudes and correlation energy for the UCCSD(4) method. The fourth-order UCCSD energy functional is given by:
\begin{equation}
\begin{aligned}
    \Delta E(4) =& \left(\langle 0| \hat{T}_{2}^{\dagger} \hat{W}_{N}|0\rangle+\text {h.c.}\right)+\langle 0| \hat{T}_{2}^{\dagger} \hat{f}_{N} \hat{T}_{2}|0\rangle+\langle 0| \hat{T}_{2}^{\dagger} \hat{W}_{N} \hat{T}_{2}|0\rangle+\langle 0| \hat{T}_{1}^{\dagger} \hat{f}_{N} \hat{T}_{1}|0\rangle \\
    &+\left(\langle 0| \hat{T}_{1}^{\dagger} \hat{W}_{N} \hat{T}_{2}|0\rangle+\text {h.c.}\right)+\frac{1}{4}\left(\langle 0| \hat{T}_{2}^{\dagger} \hat{W}_{N} \hat{T}_{2}^{2}|0\rangle+\text {h.c.}\right)
\end{aligned}
\end{equation}
Varying this functional with respect to the amplitudes yields the UCCSD(4) amplitude equations:
\begin{equation}
\begin{aligned}
    \frac{\partial \Delta E(4)}{\partial \hat{T}_1^{\dagger}} = 0 &\Rightarrow D_1 \hat{T}_1 = (\hat{W}_N \hat{T}_2)_C\\
    \frac{\partial \Delta E(4)}{\partial \hat{T}_2^{\dagger}} = 0 &\Rightarrow D_2 \hat{T}_2 = \left( \hat{W}_N + \hat{W}_N \hat{T}_2 + \hat{W}_N \hat{T}_1 + \frac{1}{4} \hat{W}_N \hat{T}_2^2 +\frac{1}{2} \hat{T}_2^{\dagger} \hat{W}_N \hat{T}_2 \right)_C,
\end{aligned}
\end{equation}
where the subscript $C$ denotes connected diagrams and $D_n$ are the orbital energy denominators. At convergence, the final UCCSD(4) energy can be calculated using a simplified expression:
\begin{equation}
    \Delta E(4)=\langle 0| \hat{W}_{N} \hat{T}_{2}|0\rangle-\frac{1}{4}\langle 0|\left(\hat{T}_{2}^{\dagger}\right)^{2} \hat{W}_{N} \hat{T}_{2}|0\rangle. 
\end{equation}

This truncation to UCCSD(4) strikes a balance between capturing essential correlation effects and computational tractability, as the inclusion of triples in the full UCC(4) method introduces additional terms that scale steeply with system size, presenting a bottleneck for larger molecules. By restricting the cluster operator to singles and doubles, UCCSD(4) preserves extensivity of the unitary formulation while enabling efficient classical implementations, making it well-suited for integration with active space approximations to model strongly correlated systems and reactions at reduced cost. However, there are potential pitfalls. If the chosen orbitals do not lead to a well-partitioned Hamiltonian---e.g., if the perturbation operator remains large or if the reference determinant poorly approximates the true ground state---the truncated expansions may converge slowly or even fail to produce reliable results.

\subsection{The Active Space UCCSD(4)/MP2 Approach}

A promising direction to reduce the computational cost of UCC calculations is to use an active space approximation, wherein UCC is applied to a carefully chosen subset of important orbitals (an active space), while treating the remaining orbitals using a cheaper method. By selecting an active space that captures the most important correlation effects--typically based on criteria such as natural orbital occupations--one can substantially reduce the cost of UCC methods without sacrificing significant accuracy, especially for large, or strongly correlated systems.

One such hybrid approach combines the accuracy of UCCSD(4) within the active space with the computational efficiency of second-order many-body perturbation theory (M) for the remaining, inactive orbitals. This hybrid framework extends the CCSD/MP2 partitioning logic of Lange and Berkelbach~\cite{Lange_Berkelbach_2020} to the UCCSD(4) method. Specifically, we consider the UCCSD(4) method and partition its cluster operator into internal (active) and external (inactive) parts:
\begin{equation}
    \hat{T} = \hat{T}_{int} + \hat{T}_{ext},
\end{equation}
where $\hat{T}_{int}$ acts only within the chosen active space, while $\hat{T}_{ext}$ accounts for excitations that involve inactive orbitals.

To reduce the computational burden, the external excitation amplitudes of $\hat{T}_{ext}$ are approximated at the M level rather than determined by iteratively solving the UCC equations. Formally, the external single and double excitations are:
\begin{equation}
\begin{aligned}
    ({\hat{T}_{1}})_{\mathrm{ext}}&=\sum_{\mathrm{ext}} t_{a}^{i} \hat{a}_a^{\dagger} \hat{a}_i \\
    ({T_{2}})_{\mathrm{ext}}&=\frac{1}{4} \sum_{\mathrm{ext}} t_{a b}^{i j} \hat{a}_a^{\dagger} \hat{a}_a^{\dagger} \hat{a}_j \hat{a}_i,  
\end{aligned}
\end{equation}
with their amplitudes given by standard MP2-like expressions:
\begin{equation}
\begin{aligned}
    t_{a}^{i} &= 0 \\
    t_{a b}^{i j}&=\frac{\langle a b \| i j\rangle}{\epsilon_i+\epsilon_j-\epsilon_a-\epsilon_b},
\end{aligned}
\end{equation}
where the $\epsilon$ are molecular orbital energies, and $\langle a b || i j\rangle$ are the anti-symmetrized two-electron integrals.

The internal excitation operator $\hat{T}_{int}$,
\begin{equation}
    \hat{T}_{\mathrm{int}}=\sum_{\mathrm{int}} t_a^i \hat{a}_a^{\dagger} \hat{a}_i+\frac{1}{4} \sum_{\mathrm{int}} t_{a b}^{i j} \hat{a}_a^{\dagger} \hat{a}_b^{\dagger} \hat{a}_j \hat{a}_i,
\end{equation}
is obtained by iteratively solving the UCCSD(4) amplitude equations restricted to the active space.

We consider two variants:
\begin{enumerate}
\item In the \textbf{composite (c-UCCSD(4)/MP2) method}, internal and external excitations are treated separately without interactive coupling of internal and external amplitudes. In the composite (c-UCCSD(4)/MP2) method, we sum separate contributions from UCCSD(4) in the active space and MP2 in the inactive space. The correlation energy $E^{comp}$ is calculated as:
\begin{equation}
    E^{\text {comp }}=E_{\mathrm{UCCSD(4)}}^{\text {act }}+ \left(E_{\mathrm{MP2}}^{\text {full }}-E_{\mathrm{MP2}}^{\text {act }}\right),
\end{equation}
where $E_{\mathrm{UCCSD(4)}}^{\text {act }}$ is the correlation energy calculated within the active space using UCCSD(4), and $E_{\mathrm{MP2}}^{\text {full}}$ and $E_{\mathrm{MP2}}^{\text {act }}$ are the MP2 correlation energies over the entire orbital and active spaces, respectively. This combination leverages the accuracy of UCCSD(4) for detailed correlations within the active space while using MP2 to efficiently account for correlations across both the active and inactive spaces.

\item Alternatively, in the \textbf{interacting (i-UCCSD(4)/MP2) method}, the internal and external amplitudes are coupled, allowing for feedback and potentially more accurate energies than the composite method. The correlation energy for the i-UCCSD(4)/MP2 method is given by:
\begin{equation}
    E^{\mathrm{int}}=\left(\sum_{\hat{T} \in \mathrm{int}}+\sum_{\hat{T} \in \mathrm{ext}}\right) E_{UCCSD(4)}\left[ \hat{T} \right]
\end{equation}
\end{enumerate}

Both variants offer reduced scaling compared to full UCCSD(4), with the choice depending on the desired balance between accuracy and computational efficiency. 

To investigate the impact of orbital choice on the performance of our active space UCCSD(4)/MP2 methods, we employ both canonical orbitals (COs) and frozen natural orbitals (FNOs)~\cite{sosa_selection_1989}. Canonical orbitals are the eigenfunctions of the Fock operator, providing a delocalized basis which is optimal for the mean-field description and often leads to stable performance in post-HF methods. The FNO approach exploits the fact that a significant portion of dynamic correlation is captured by a relatively small subset of virtual orbitals with the largest occupation numbers. By truncating the virtual space, the computational cost of post-HF methods, whose most expensive steps scale steeply with the number of virtual orbitals, is substantially reduced. In this work, FNOs are generated by constructing the virtual–virtual block of the unrelaxed MP2 one-particle reduced density matrix (1-RDM) and diagonalizing it to obtain natural virtual orbitals ranked by their occupation numbers. A subset of virtual orbitals with the largest MP2 occupations is retained in the active space, while the remaining virtual orbitals are frozen and treated perturbatively at the MP2 level. The occupied orbitals are kept canonical in all cases. Unless stated otherwise, all references to natural orbitals (NOs) in this work correspond to this FNO construction. Comparing CO- and NO-based implementations allows us to assess the trade-offs between numerical robustness and orbital compactness within both the composite and interacting active-space frameworks.

\section{Computational Details}

Geometry optimizations and Nudged Elastic Band (NEB) calculations~\cite{neb_paper} for reaction paths were carried out with the ORCA program~\cite{orca, orca6}. Selected Configuration Interaction with Perturbation Theory (CIPSI) calculations were performed using the Quantum Package~\cite{quantum_package}. All other electronic structure calculations were performed using the PySCF package~\cite{sun_pyscf_2018}. The one- and two-electron integrals required for our custom post-Hartree-Fock implementations were also computed using PySCF. Frozen Natural Orbitals (FNOs) were generated from MP2 calculations using PySCF by diagonalizing the virtual–virtual block of the MP2 one-particle reduced density matrix. Virtual orbitals were ranked by occupation number, with the most occupied retained in the active space and the remainder frozen and treated at the MP2 level. The symbolic algebraic expressions for the UCCSD(4) amplitude and energy equations were derived using the pdaggerq~\cite{rubin_pq_2021} and Wick\&d~\cite{evangelista_automatic_2022} packages, which facilitate the automated derivation and implementation of many-body theories. The custom implementation of the active space $UCCSD(4)/MP2$ methods, including the composite and interacting variants, is available on GitHub at \nolinkurl{https://github.com/prateekvaish/est.git}. Specific implementation details, such as basis sets and active space selections, are provided in the respective Results sections.

\section{Results and Discussion}

\subsection{Performance of Active Space UCC on Small Molecules in Their Equilibrium Geometries}

\begin{figure}[ht]
    \centering
    \begin{subfigure}[t]{0.49\linewidth}
        \centering
        \includegraphics[width=\linewidth]{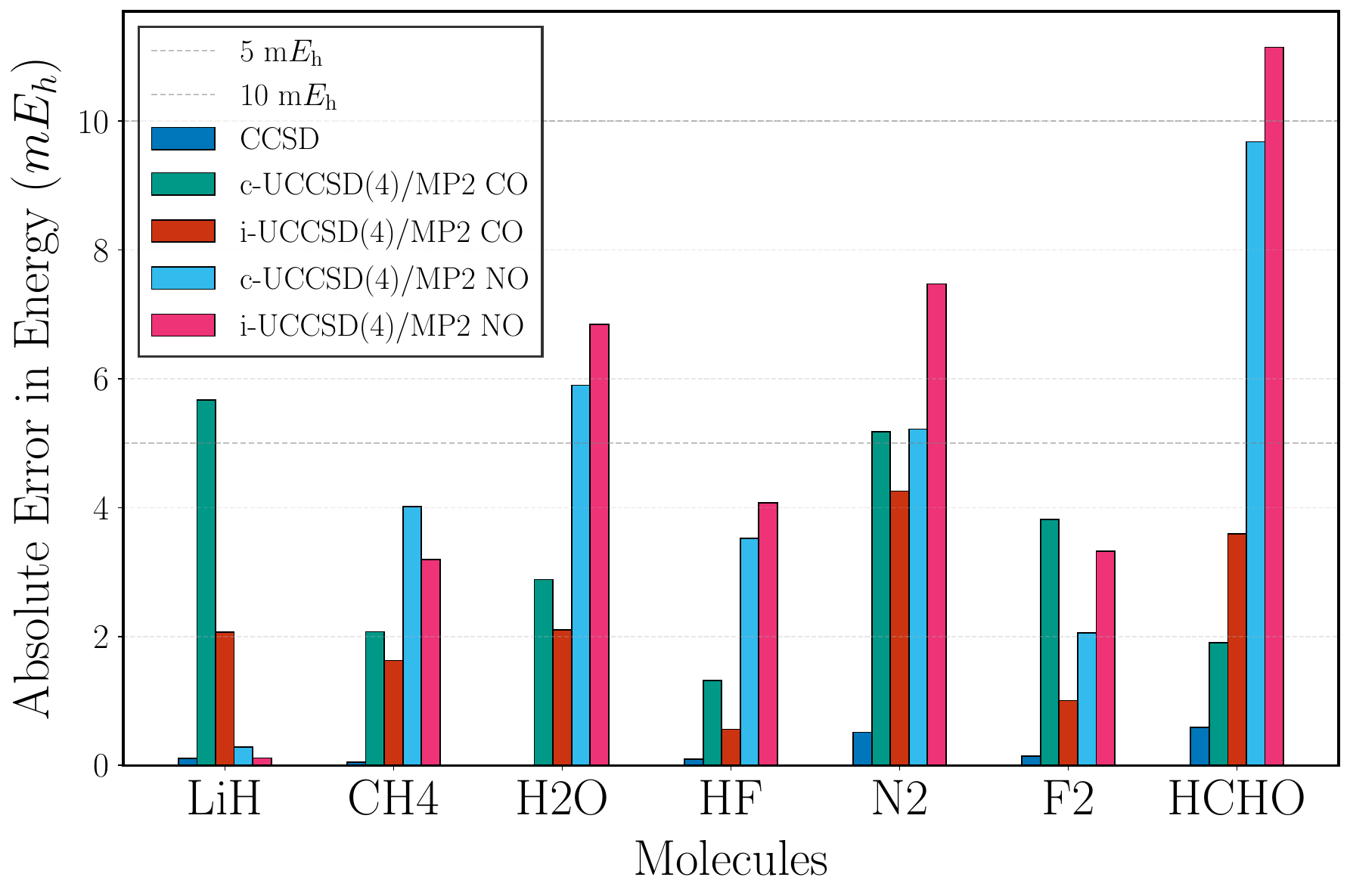}
        \caption{cc-pVDZ basis}
        \label{fig: dz equilibrium dataset}
    \end{subfigure}
    \hfill
    \begin{subfigure}[t]{0.49\linewidth}
        \centering
        \includegraphics[width=\linewidth]{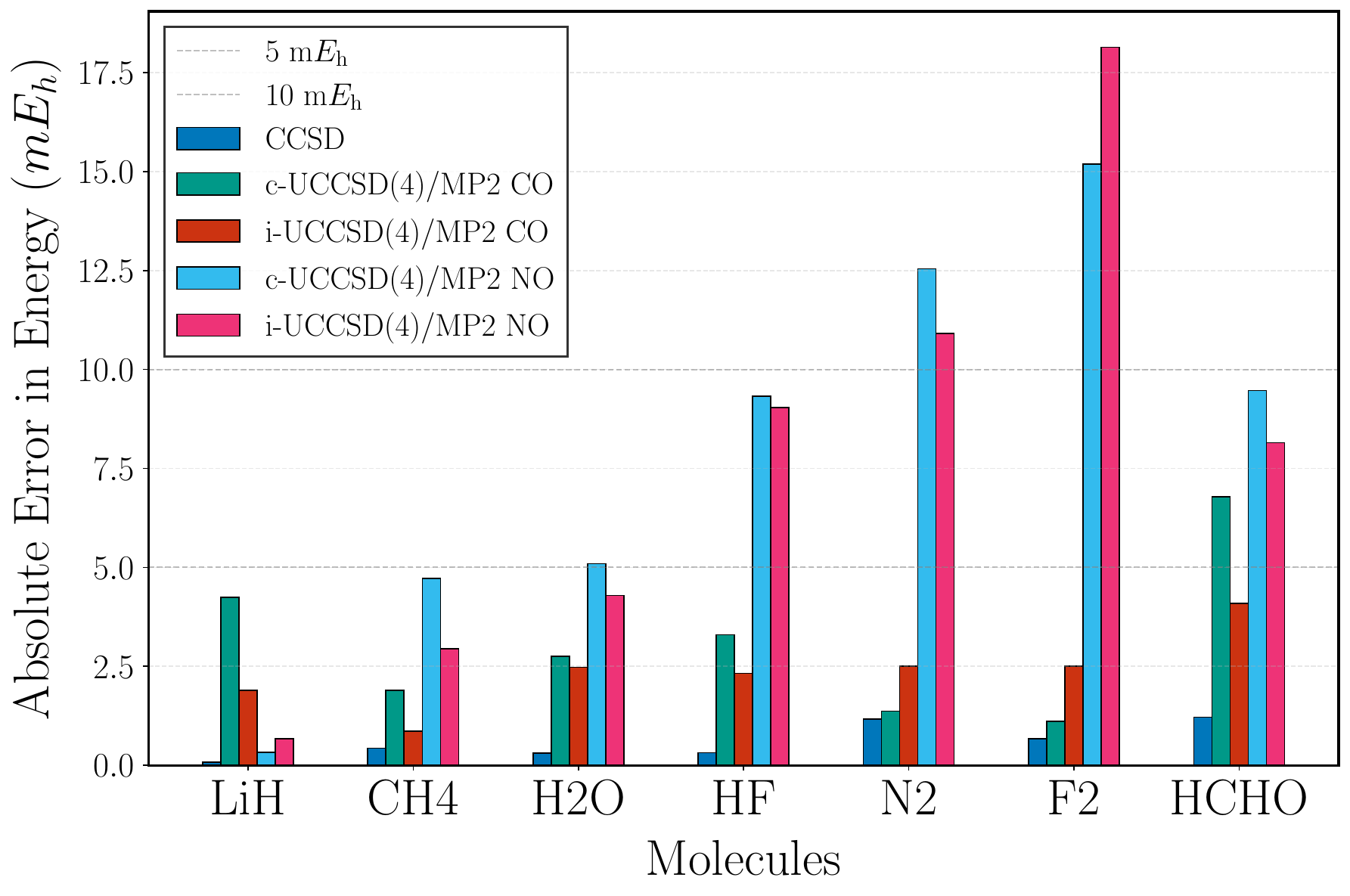}
        \caption{cc-pVTZ basis}
        \label{fig: tz equilibrium dataset}
    \end{subfigure}
    \caption{Absolute errors in total energies (in milliHartrees) relative to the UCCSD(4) reference for a range of molecules. Each bar represents a different approximation: CCSD (blue); composite UCCSD (c-UCCSD(4)/MP2) with canonical orbitals (CO) (dark teal); 
    interacting UCCSD (i-UCCSD(4)/MP2) with canonical orbitals (red); composite UCCSD (c-UCCSD(4)/MP2) with natural orbitals (NO) (cyan); and
    interacting UCCSD (i-UCCSD(4)/MP2) with natural orbitals (magenta). Only 60\% of the virtual orbitals are retained in the active space.}
    \label{fig: equilibrium_dataset_combined}
\end{figure}

We begin by assessing the performance of our active space UCCSD(4)/MP2 methods on the total energies of several small molecules at their equilibrium geometries using the cc-pVDZ and cc-pVTZ basis sets. These molecules--LiH, CH$_4$, H$_2$O, HF, N$_2$, F$_2$, and H$_2$CO--exhibit weak to moderate electron correlation, as indicated by low T1 diagnostic values ($<$0.02) and thus serve as initial benchmarks to evaluate the accuracy of the approximations in regimes where the single-reference Hartree-Fock wavefunction is reasonable. The equilibrium geometries were obtained from the Computational Chemistry Comparison and Benchmark Database (CCCBDB)~\cite{NIST_CCCBDB_2022}. Unless otherwise stated, all accuracy assessments in this section are made relative to full UCCSD(4), which serves as our internal reference for evaluating active space approximations; comparisons to CCSD(T) are used only to contextualize chemical accuracy.

In these calculations, the active space includes all occupied valence orbitals and 60\% of the virtual orbitals. Virtual orbitals are chosen based on MP2 natural orbital occupations to prioritize those with the largest contributions to the correlation. Figures ~\ref{fig: dz equilibrium dataset} and ~\ref{fig: tz equilibrium dataset} illustrate the absolute errors in total energies (in milliHartrees) relative to UCCSD(4) for CCSD and the two active space UCCSD(4)/MP2 variants. 

In the cc-pVDZ basis (Figure \ref{fig: dz equilibrium dataset}), CCSD agrees with UCCSD(4) to within 1 mHa for all molecules, with a mean absolute error (MAE) of 0.2 mHa and a maximum error of 0.6 mHa (for CH$_2$O). This close agreement reflects the similarity between traditional CCSD and UCCSD(4) in weakly correlated systems. The active space methods exhibit larger errors but remain chemically accurate for many purposes, with MAEs ranging from 2.2 mHa (i-UCCSD(4)/MP2 CO) to 5.2 mHa (i-UCCSD(4)/MP2 NO). 

A key observation is that canonical orbitals (COs) generally lead to lower absolute errors relative to UCCSD(4). This behavior can be attributed to the fact that canonical orbitals preserve a strictly diagonal zeroth-order Hamiltonian across the full orbital space, ensuring well-defined energy denominators and a clean many-body perturbation theory (MBPT) partitioning underlying the UCCSD(4) truncation. However, the choice of orbital basis has nuanced effects when considering errors relative to CCSD(T), where the composite method with natural orbitals (c-UCCSD(4)/MP2 NO) yields lower errors than its CO counterpart. This is in agreement with the previous study by Berkelbach and coworkers,~\cite{Lange_Berkelbach_2020} in which they found that NOs provide better energies for the CCSD/MP2 method when benchmarked against CCSD(T). This improvement arises from the more compact representation of correlation provided by NOs, allowing a truncated virtual space to recover a larger fraction of the dynamic correlation energy. 

In contrast, the interacting NO variant consistently performs worse than its CO counterpart. Importantly, this behavior should not be interpreted as an intrinsic deficiency of natural orbitals. Rather, it reflects an incompatibility between the interacting UCCSD(4)/MP2 formalism and the orbital structure induced by frozen natural orbital (FNO) truncation. The UCCSD(4) method employed here is formulated assuming canonical orbitals. While variants exist that utilize semi-canonical orbitals, such as those proposed by Windom et al.~\cite{Windom_Claudino_Bartlett_2024}, semi-canonicalization requires the occupied-occupied and virtual-virtual blocks of the Fock matrix to be diagonal, under the present NO scheme (Frozen Natural Orbitals), semi-canonicalization allows for diagonal occupied-occupied and virtual-virtual blocks within the active space; however, the virtual-virtual block outside the active space is no longer diagonal. As a result, when the interacting UCCSD(4)/MP2 energy expression is evaluated over the full orbital space, the implicit MBPT ordering assumed in the fourth-order truncation is violated. This allows higher-order external contributions to contaminate nominal fourth-order terms, leading to numerical instability and systematically larger errors. This issue does not arise in the composite method, where the UCCSD(4) equations are applied exclusively within the active space for which the Fock operator remains diagonal.

In the larger cc-pVTZ basis (Figure \ref{fig: tz equilibrium dataset}), errors increase overall due to the expanded virtual space (approximately twice as many virtual orbitals as in cc-pVDZ). CCSD maintains good agreement with UCCSD(4), with an MAE of 0.6 mHa and a maximum error of 1.2 mHa (for H$_2$CO). The active space methods show MAEs from 2.4 mHa (i-UCCSD(4)/MP2 CO) to 7.7 mHa (i-UCCSD(4)/MP2 NO). This increase reflects not only the larger virtual space in cc-pVTZ, but also the fixed 60\% truncation criterion, which corresponds to a substantially larger number of discarded virtual orbitals compared to cc-pVDZ. The i-UCCSD(4)/MP2 NO variant exhibits particularly large errors, exceeding 10 mHa for $N_2$ and reaching approximately 18 mHa for $F_2$, further underscoring the sensitivity of the interacting formulation to orbital non-canonicality.

Overall, these benchmarks demonstrate that our active space UCCSD(4)/MP2 methods can achieve reliable accuracies with significantly reduced computational demands. Canonical orbitals emerge as the default and most robust choice, particularly for the interacting formulation, due to their consistency with the perturbative structure of UCCSD(4). Natural orbitals, while advantageous for compactness and beneficial in composite schemes, can introduce instabilities when combined with the interacting UCCSD(4)/MP2 approach. Consequently, the interacting method is best suited for weakly to moderately correlated systems when orbital semi-canonicality is preserved across the full space, whereas the composite approach offers a simpler and more robust alternative when aggressive orbital truncation is employed.

\subsection{Performance on the GW100 Dataset}

\begin{figure}[ht]
    \centering
    \begin{subfigure}[t]{0.49\linewidth}
        \centering
        \includegraphics[width=\linewidth]{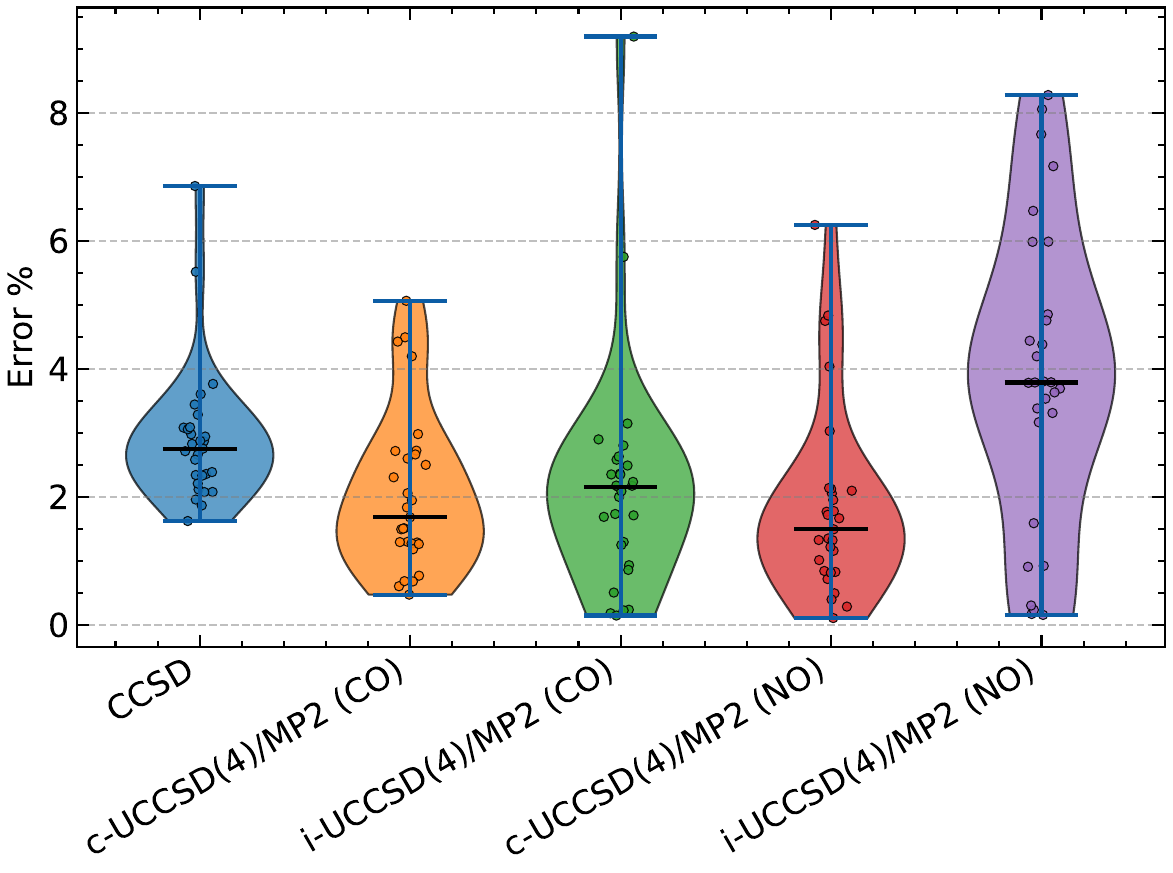}
        \caption{Medium-sized molecule benchmark set}
        \label{fig: medium_pct_violin}
    \end{subfigure}
    \hfill
    \begin{subfigure}[t]{0.49\linewidth}
        \centering
        \includegraphics[width=\linewidth]{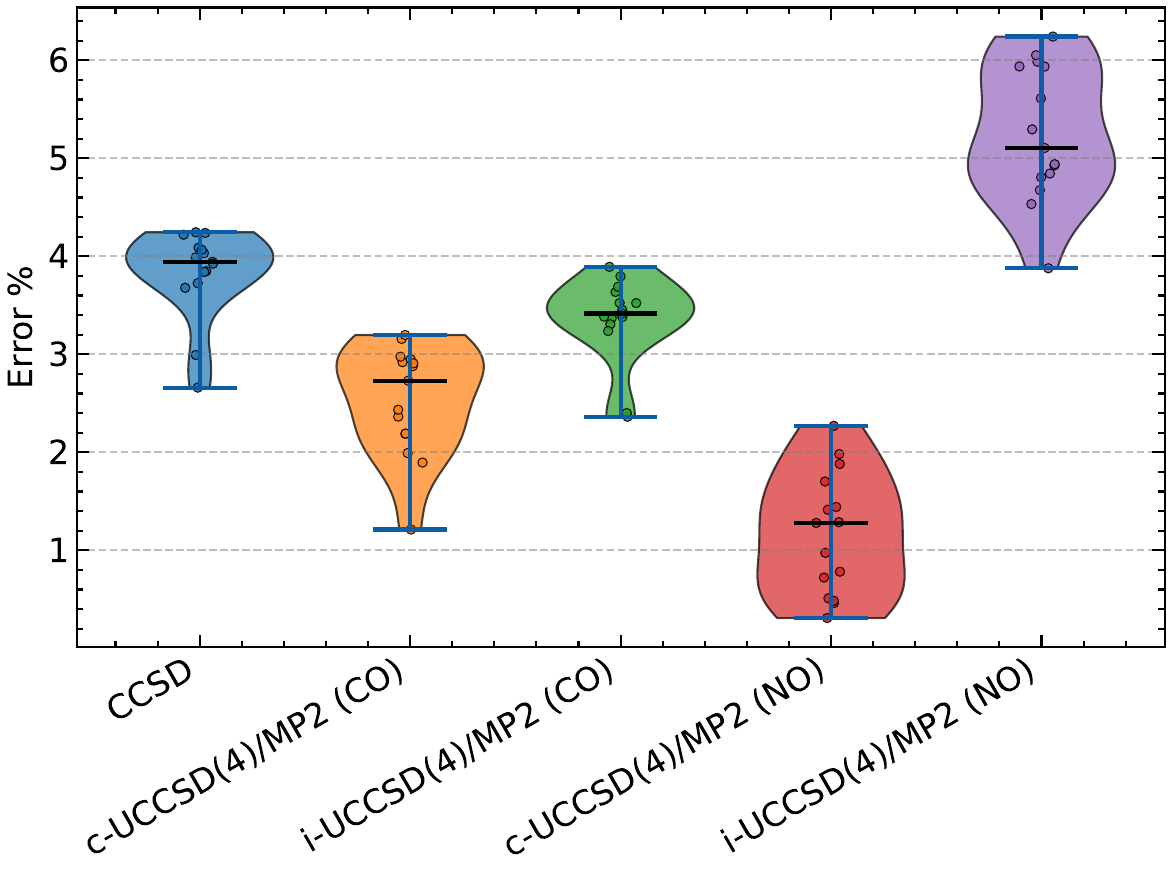}
        \caption{Large molecule benchmark set}
        \label{fig: large_pct_violin}
    \end{subfigure}
    \caption{Percentage error in correlation energies with respect to the CCSD(T) reference for (a) medium‐sized and (b) large molecules from the GW100 dataset. Each violin envelope shows the distribution of errors over all molecules in the set; the thick horizontal bar marks the median and the whiskers span the full range, while the overlaid points correspond to individual molecules. ``CCSD'' is the full canonical CCSD result.  All other methods employ the active–space \mbox{UCCSD(4)/MP2} method. ``CO'' and ``NO'' indicate that the calculations were performed using canonical and natural orbitals, respectively.}
    \label{}
\end{figure}

We next assess the performance of our active space methods across subsets of the GW100 dataset~\cite{van_setten_gw100_2015}, comprising 29 medium-sized molecules (e.g., silane, methane, alcohols, and halides) and 15 large molecules (e.g., benzene derivatives and nucleobases). The GW100 dataset includes diverse organic and inorganic species, providing a broader test of active space approximations. For this benchmark, we use the highly accurate CCSD(T) method as the reference and focus on the error in the correlation energy. All calculations reported in this section were performed using the cc-pVDZ basis set. The active space includes all occupied valence orbitals and 50\% of the virtual orbitals, selected based on MP2 natural orbital occupations. Figures \ref{fig: medium_pct_violin} and \ref{fig: large_pct_violin} show the distribution of the percentage errors in the correlation energy for the medium and large molecule sets, respectively. These plots capture the distribution, medians, and outliers, offering insights into overall performance. Absolute errors in milliHartrees (mEh) and detailed breakdowns per molecule represented as heatmaps are available in the Supplemental Information.

For the medium-sized molecules (Figure ~\ref{fig: medium_pct_violin}), CCSD recovers most of the correlation energy, with a median percentage error of approximately 2.8\%. The active-space methods, while more approximate, capture a substantial fraction of the correlation energy, with median errors clustering between roughly 1.6\% and 4\%. Notably, when benchmarked against CCSD(T), the composite method using natural orbitals (c-UCCSD(4)/MP2 NO) yields the lowest median error (1.5\%), outperforming its canonical-orbital counterpart (c-CO, median 1.7\%). This behavior is consistent with previous observations that natural orbitals provide a more compact representation of correlation, enabling a truncated virtual space to recover a larger fraction of the correlation energy. In contrast, the interacting NO variant consistently performs worse than its canonical orbital counterpart, with a median error of approximately 3.8\% compared to about 2.2\% for the corresponding CO variant. This performance gap arises because the interacting UCCSD(4) energy expression is sensitive to the non-diagonal virtual–virtual blocks of the Fock matrix that remain in the external space under the frozen-natural-orbital scheme, leading to reduced numerical stability and systematically larger errors.

A similar overall trend is observed for the large-molecule subset (Figure \ref{fig: large_pct_violin}). Interestingly, the percentage error distributions are noticeably tighter for the large molecules than for the medium-sized set. This reflects the fact that the total correlation energy grows substantially with system size, causing absolute errors, which do increase with size, to represent a smaller fraction of the total correlation energy. Within this set, the composite NO method again performs best, achieving a median error of approximately 1.3\%, while the c-CO and i-CO variants cluster around median errors of roughly 2.7\% and 3.4\%, respectively. The interacting NO variant remains the least accurate, with a median error of approximately 5.1\% and a broader error distribution.

These results confirm that the active space approximation captures a consistent and significant portion of the correlation energy, even as the system size increases. Moreover, while natural orbitals offer superior compactness and improved accuracy for the composite active-space approach, canonical orbitals remain the more robust choice for the interacting UCCSD(4)/MP2 framework. The composite scheme often provides the best balance between accuracy and efficiency across chemically diverse systems by exploiting the compactness of NOs without incurring the instability inherent to the interacting formulation.

\subsection{Phosphate Hydrolysis: Performance on a Moderately-Correlated Reaction}

\begin{figure}[h!]
    \centering
    \includegraphics[width=\linewidth]{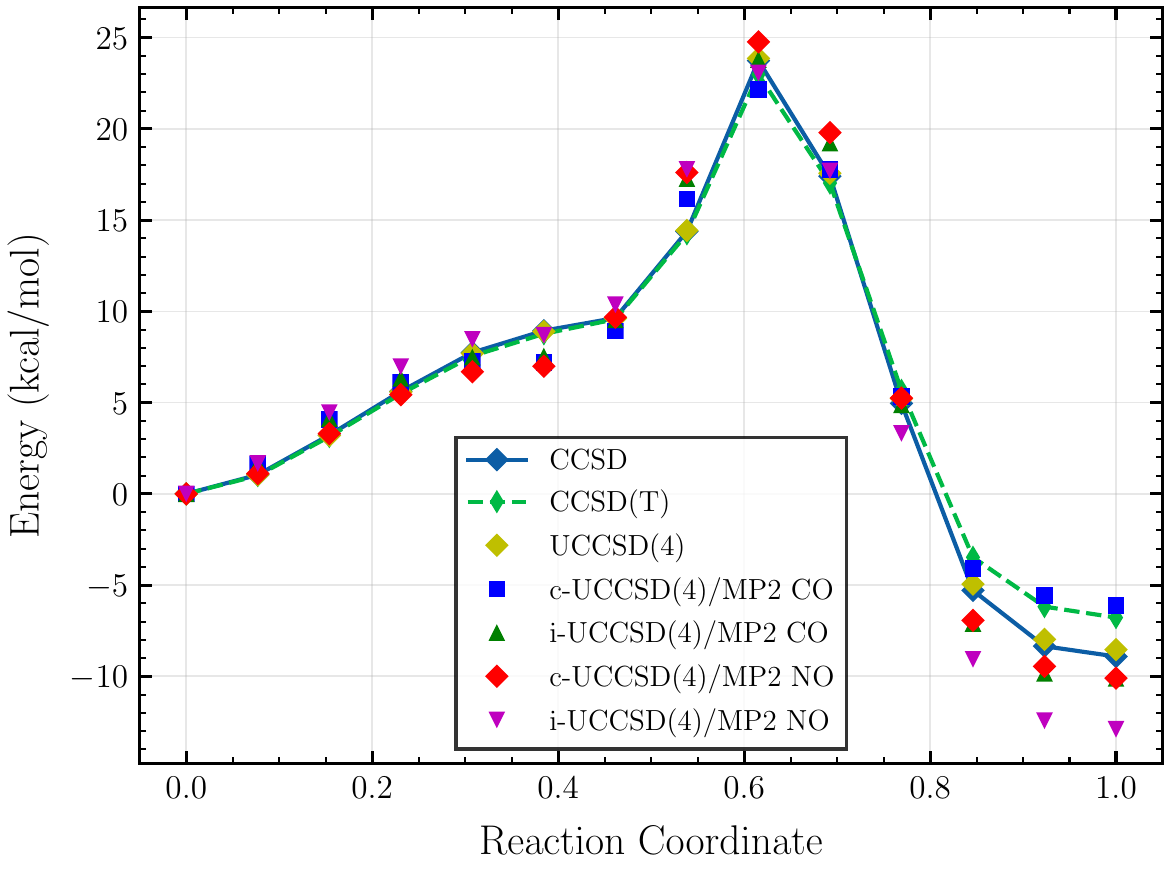}
    \caption{Reaction Energy Profile for the PO$_3^-$ + H$_2$O reaction, computed using the cc-pVDZ basis set. Electronic energies are reported relative to the reactant energy. The CCSD(T) curve (green, dashed line) serves as the high-level reference, while the blue trace (solid line) and yellow squares show the corresponding CCSD and UCCSD(4) results. All remaining curves were obtained with the active-space \mbox{UCCSD(4)/MP2} methods, where the active space contains all occupied valence orbitals and 11 virtual orbitals. ``i-'' denotes the interacting UCCSD(4)/MP2 method, ``c-'' the composite UCCSD(4)/MP2 method, and ``CO'' and ``NO'' refer to canonical orbitals and natural orbitals, respectively.}
    \label{fig:reaction_profile}
\end{figure}

\begin{table}
    \centering
    \caption{Barrier heights (in kcal/mol) of the PO$_3^-$ + H$_2$O reaction. The active space for the UCCSD(4)/MP2 methods contain all occupied valence orbitals and 11 virtual orbitals.}
    \label{tab:combined_barrier_tables}
    \begin{tabular}{l r r}
        \toprule
        Method & Forward $\Delta E^\ddagger$ & Backward $\Delta E^\dagger$ \\
        \midrule
        CCSD & 23.7 & 32.7 \\
        UCCSD(4) & 23.9 & 32.4 \\
        CCSD(T) & 23.0 & 29.8 \\
        \midrule
        c-UCCSD(4)/MP2 CO & 22.2 & 28.3 \\
        i-UCCSD(4)/MP2 CO & 23.8 & 33.9 \\
        c-UCCSD(4)/MP2 NO & 24.8 & 34.9 \\
        i-UCCSD(4)/MP2 NO & 23.1 & 36.0 \\
        \bottomrule
    \end{tabular}
\end{table}

\begin{figure}[ht]
    \centering
    \includegraphics[width=\textwidth]{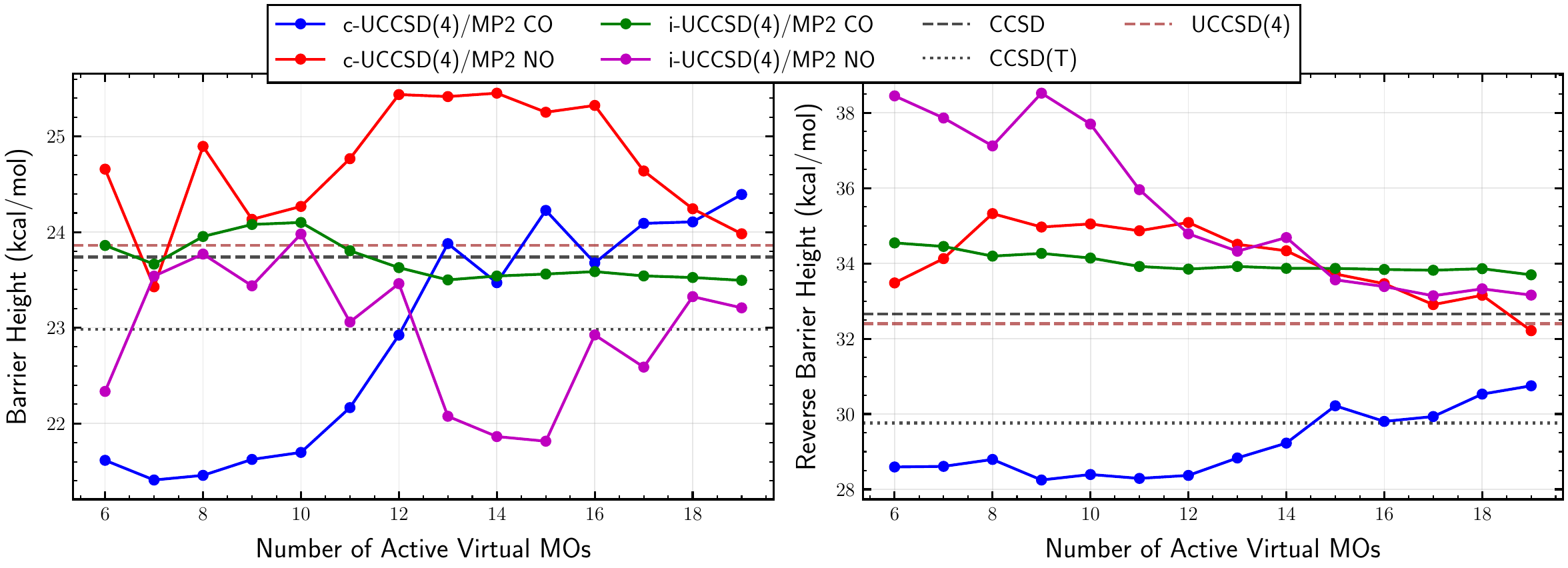}
    \caption{Dependence of the barrier height for the forward (left panel) and backward (right panel) PO$_3^-$ + H$_2$O reaction on the number of virtual orbitals included in the active space. All calculations were performed using the cc-pVDZ basis set. The active space for these calculations contained all occupied valence orbitals. The full-space CCSD, UCCSD(4), and CCSD(T) values are provided as horizontal lines for reference. The ``i-'' and ``c-'' prefixes denote the interacting and composite UCCSD(4)/MP2 methods, respectively, while ``CO'' and ``NO'' refer to the use of canonical and natural orbitals.}
    \label{fig:combined_barriers}
\end{figure}

Beyond benchmarking on molecules in their equilibrium geometries, we now evaluate the performance of our active space methods on a chemical reaction: the hydrolysis of the metaphosphate anion (PO$_3^-$) with water. This reaction serves as a minimal model for ATP hydrolysis, a fundamental process in biochemistry~\cite{plotnikov_quantifying_2013}. Its reaction profile, involving bond formation and cleavage at the transition state, presents a more rigorous test for assessing the balanced description of static and dynamic electron correlation. All calculations were performed using the cc-pVDZ basis set.

Figure \ref{fig:reaction_profile} illustrates the reaction energy profile calculated with various methods, using an active space containing all occupied valence orbitals and 11 virtual orbitals. Overall, the active space UCCSD(4)/MP2 methods closely follow the qualitative shape of the full space CCSD and UCCSD(4) potential energy curves along the reaction coordinate. The full UCCSD(4) slightly overestimates the barrier compared to CCSD(T). Among the active space approximations, the interacting variants reproduce the UCCSD(4) profile more faithfully near the transition state than the composite methods, whereas the composite approaches exhibit larger deviations in the barrier region. For both schemes, calculations performed using canonical orbitals yield smoother and more systematic behavior than those using natural orbitals.

Barrier heights for the forward and reverse reactions are summarized in Table~\ref{tab:combined_barrier_tables}. For the forward reaction, CCSD and UCCSD(4) yield similar barrier heights, both lying above the CCSD(T) reference. The active space methods span a wider range of values. The interacting canonical orbital variant closely reproduces the UCCSD(4) forward barrier, whereas the composite canonical approach underestimates it. In contrast, the natural orbital variants show larger deviations. The composite NO method overestimates the forward barrier, while the interacting NO result lies closer to CCSD(T). This behavior highlights the competing effects of orbital compactness and coupling sensitivity in determining barrier heights. 

A similar pattern is observed for the reverse barrier. CCSD and UCCSD(4) again predict larger barriers than CCSD(T). The interacting canonical orbital method remains the closest approximation to the UCCSD(4) reference, while the composite canonical approach significantly underestimates the reverse barrier. Both NO-based variants overestimate the reverse barrier, with the interacting NO method showing the largest deviation. Taken together, these results indicate that the interacting scheme is more effective at preserving the quantitative features of the full UCCSD(4) method, while the composite approach can benefit from partial error cancellation when compared directly to CCSD(T).

To assess the robustness of these conclusions, we examine the dependence of the predicted barrier heights on the size of the active virtual space. Figure~\ref{fig:combined_barriers} shows the forward and reverse barrier heights as a function of the number of virtual orbitals included in the active space. When canonical orbitals are employed, the interacting method exhibits smooth and monotonic convergence as the active space is enlarged when canonical orbitals are employed, with only modest fluctuations across the entire range of virtual orbital counts. In contrast, the interacting natural orbital variant displays larger residual oscillations, indicating increased sensitivity to the active space definition despite retaining the overall qualitative trend. The composite methods show oscillations that persist even as additional virtual orbitals are included. This sensitivity suggests that the composite formulation is less stable with respect to active space definition for this reaction. Canonical orbitals consistently converge faster and with smaller fluctuations than natural orbitals for both interacting and composite variants. In particular, the interacting canonical orbital method reaches near-converged barrier heights with a relatively small fraction of the virtual space, whereas the NO-based interacting variant shows larger residual variations even at the largest active spaces considered.

Overall, these results demonstrate that the active-space UCCSD(4)/MP2 framework is capable of describing moderately correlated reaction profiles with good qualitative fidelity and in favorable cases, quantitative accuracy. The interacting formulation provides the most reliable approximation to full UCCSD(4) along the reaction coordinate and exhibits superior stability with respect to active space size, especially when canonical orbitals are employed. The composite approach, while computationally simpler, is more sensitive to the choice of active space and orbital basis, though it can occasionally benefit from fortuitous error cancellation when compared against CCSD(T).

\subsection{Ethylene Torsion: A Test for Static Correlation}

\begin{figure}[h!]
    \centering
    \includegraphics[width=0.9\linewidth]{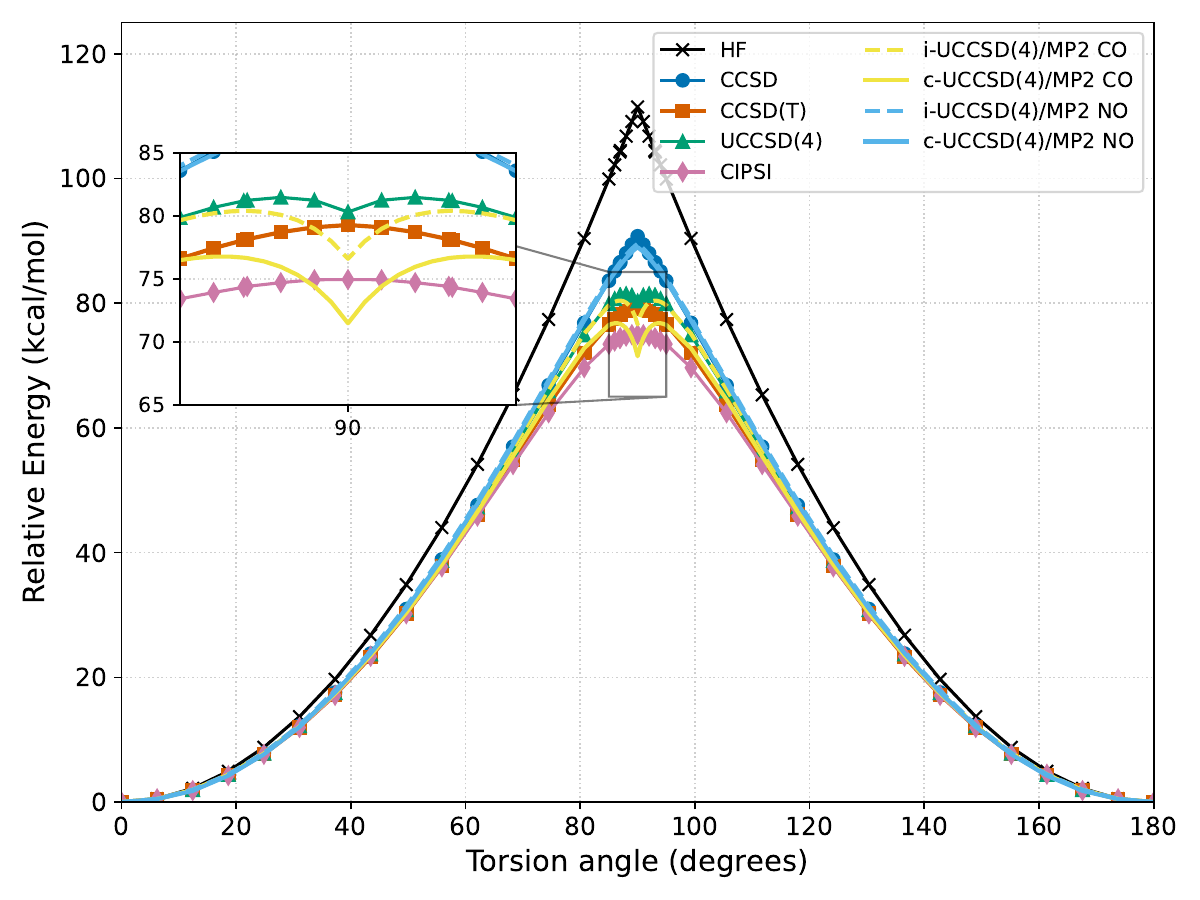}
    \caption{Potential energy curve for the torsion of the ethylene molecule. Electronic energies are reported in kcal/mol relative to the planar structure (0$^\circ$ torsion angle). The plot compares results from several electronic structure methods. The CCSD(T) curve (orange squares) and the CIPSI curve (pink diamonds) serve as high-level references. The corresponding results for HF, CCSD, and UCCSD(4) are also shown. The remaining curves were obtained with active-space UCCSD(4)/MP2 methods,  where the active space contains all occupied valence orbitals and 22 virtual orbitals. ``i'' denotes the interacting method, ``c-'' denotes the composite method, and ``CO'' and ``NO'' refer to the use of canonical orbitals and natural orbitals, respectively.}
    \label{fig:ethylene_plot}
\end{figure}

\begin{table}[h!]
    \centering
    \begin{tabular}{l r}
        \toprule
        Method & $\Delta E$ [in kcal/mol]\\
        \midrule
        RHF & 111.5 \\
        MP2 & 102.6 \\
        CCSD & 90.7 \\
        CCSD(T) & 79.3 \\
        UCCSD(4) & 81.5 \\
        CIPSI & 75.0 \\
        \midrule
        c-UCCSD(4)/MP2 CO & 76.8 \\
        i-UCCSD(4)/MP2 CO & 80.4 \\
        c-UCCSD(4)/MP2 NO & 89.0 \\
        i-UCCSD(4)/MP2 NO & 89.5 \\
        \bottomrule
    \end{tabular}
    \caption{Torsional Barrier(in kcal/mol) of Ethylene. The active space for the UCCSD(4)/MP2 methods contain all occupied valence orbitals and 22 virtual orbitals}
    \label{tab:ethylene_torsion}
\end{table}

The torsional rotation of ethylene around its C--C bond is a classic benchmark for electronic structure methods, as it involves a transition from a closed-shell ground state at 0$^\circ$ (planar) to a diradical state at 90$^\circ$ (perpendicular), where static correlation is dominant~\cite{ben-nun_photodynamics_2000}. In the twisted configuration, the near-degeneracy of the HOMO ($\pi$) and LUMO ($\pi^*$) orbitals gives rise to pronounced static correlation, rendering single-determinant references qualitatively inadequate~\cite{barbatti_photochemistry_2004}. As a result, this system provides a stringent test of whether an approximate method can remain reliable as the electronic structure departs significantly from a single-reference description.

The potential energy curves for the torsion are shown in Figure~\ref{fig:ethylene_plot} and the calculated barriers using different different methods are listed in Table~\ref{tab:ethylene_torsion}. The difficulty of describing this electronic state is highlighted by our CIPSI calculations, which necessitate a large number of determinants to achieve convergence, a finding consistent with the broader literature~\cite{barbatti_photochemistry_2004}. This highlights the multireference nature of the perpendicular geometry and the importance of capturing static correlation effects accurately. As expected, single-reference methods that do not capture static correlation adequately perform poorly. The Hartree-Fock (HF) method significantly overestimates the energy throughout the rotation. MP2 and CCSD also overestimate the barrier heights. The inclusion of perturbative triples in CCSD(T) significantly improves the result, lowering the barrier to 79.3 kcal/mol, yielding a smoother and more physically reasonable potential energy curve. For a high-accuracy benchmark, we used the CIPSI method, which predicts a barrier of 75.0 kcal/mol, indicating the importance of excited determinants for this strongly correlated system. UCCSD(4) method yields a curve similar to CCSD(T) despite the fact that it does not contain triples, reflecting its ability to capture higher-order correlations beyond the traditional CCSD method. However, a closer inspection of the potential energy curve reveals small, unphysical dips near the 90$^\circ$ transition state. These artifacts are symptomatic of the breakdown of single-reference methods in regions of strong static correlation where the electronic structure has significant multi-reference character from more than one dominant configuration. Although the Baker-Campbell-Hausdorff (BCH) truncated formalism makes UCCSD(4) theoretically more robust than traditional CCSD, it cannot entirely circumvent the deficiencies of its single-reference nature in such challenging cases.

Our active space UCCSD(4)/MP2 variants--both composite (c-) and interacting (i-), and employing either canonical orbitals(CO) or natural orbitals(NO)--exhibit diverse behaviors in this challenging regime. In contrast to their performance for weakly and moderately correlated systems, the choice of orbital basis plays a dominant role in determining the quality of the torsional profile. When canonical orbitals are employed, both the composite and interacting formulations closely track the full UCCSD(4) potential energy curve and yield torsional barriers that are in reasonable agreement with the high-level CIPSI reference. As expected for active space approximations, neither variant resolves the small unphysical features present in the full UCCSD(4) curve near the perpendicular geometry, but they faithfully reproduce its overall shape and barrier height.

With the canonical orbitals, the interacting UCCSD(4)/MP2 method provides the closest approximation to the full UCCSD(4) curve across the torsional coordinate, reflecting its tighter numerical consistency with the underlying UCCSD(4) solution within the active space while treating external correlation perturbatively. The composite formulation with canonical orbitals performs comparably, albeit with slightly larger deviations in the barrier region. In contrast, both composite and interacting variants employing natural orbitals deviate substantially from the UCCSD(4) and CIPSI references and instead yield torsional profiles that more closely resemble CCSD. This behavior indicates a breakdown of the natural orbital truncation strategy in the presence of strong static correlation, where orbital compactness does not translate into an improved representation of near-degenerate configurations. These observations underscore an important distinction between orbital choice and coupling strategy in strongly correlated systems. For the ethylene torsion, the failure of the NO-based variants dominates the error profile, while both composite and interacting formulations remain viable when canonical orbitals are used, even though the interacting approach offers a systematically closer approximation to full UCCSD(4) method.

To improve the accuracy and eliminate the unphysical artifacts observed in the UCCSD(4) curve, one could pursue more computationally demanding approaches. One avenue is the inclusion of higher-order external amplitudes within the active-space framework, providing a more complete description of dynamic correlation. Alternatively, one could systematically improve the underlying UCC formalism. One route is to include more terms in the Baker-Campbell-Hausdorff (BCH) expansion to create higher order approximations beyond the one used here. A more powerful, albeit challenging, approach would be to employ a theoretically exact, untruncated UCC formulation, a method often targeted for implementation on quantum computers. Both strategies would systematically improve the treatment of correlation and yield a more reliable potential energy surface at a substantially greater computational cost.

\section{Conclusions}
\label{sec:conclusions}
In this work, we have introduced and systematically benchmarked an active space Unitary Coupled Cluster approach, UCCSD(4)/MP2, designed to model electron correlation in chemical systems with reduced computational cost. By combining the accuracy of UCCSD(4) for a defined active space with the efficiency of MP2 calculations for the external space, we have developed a powerful hybrid method suitable for a range of chemical problems. Our findings demonstrate that this active space framework, particularly the interacting i-UCCSD(4)/MP2 variant using canonical orbitals, provides a robust and reliable approximation to the full UCCSD(4) method. When tested on small molecules at their equilibrium geometries and medium and large molecules from the GW100 dataset, the method consistently captures essential correlation effects, yielding chemically accurate results with only a fraction of the virtual orbitals. Across these benchmarks, canonical orbitals emerge as the most robust choice for the interacting formulation, whereas natural orbitals can be advantageous within the composite scheme by providing a more compact representation of dynamic correlation. 

The performance of our method on the phosphate hydrolysis reaction highlights its potential for modeling chemically relevant processes involving bond breaking and formation. The interacting i-UCCSD(4)/MP2 approach using canonical orbitals not only reproduced the UCCSD(4) reaction profile but also showed remarkable stability with respect to the active space size. This stability is a critical advantage over the composite variants, which exhibited erratic behavior, making them less predictable for chemical applications. Although composite methods can benefit from partial error cancellation when compared directly to CCSD(T), the interacting method provides a more faithful approximation to the full UCCSD(4) method along the reaction coordinate, achieving chemical accuracy with as few as 15--25\% of the virtual orbitals included in the active space.

In contrast, the ethylene torsion benchmark exposes the limitations of the active-space UCCSD(4)/MP2 framework in regimes dominated by strong static correlation. In this challenging case, both composite and interacting variants using canonical orbitals closely track the full UCCSD(4) potential energy curve and yield torsional barriers in reasonable agreement with high-level references, but they do not eliminate the unphysical features inherited from the single-reference UCCSD(4) ansatz near the perpendicular geometry. Variants employing natural orbitals perform significantly worse and revert toward CCSD-like behavior. These results highlight that the relative performance of the composite and interacting formulations depends sensitively on both the orbital representation and the nature of the correlation. 

The primary motivation for this active space approach is the substantial reduction in computational cost. Full UCCSD(4), like CCSD, has a computational scaling that is formally $O(o^4 v^2 + o^3 v^3 + o^2 v^4)$, where $o$ is the number of occupied orbitals and $v$ is the number of virtual orbitals. This leads to an effective worst-case scaling of $O(N^6)$, assuming $o \approx v \approx N/2$ (where N is the total number of basis functions, which typically serves as a measure of system size). The most computationally intensive step typically scales as $O(o^2 v^4)$, especially since $v >> o$ in large basis sets. In our UCCSD(4)/MP2 framework, the expensive iterative UCCSD(4) calculation is confined to the active space, reducing its cost to $O(o_{act}^2 v_{act}^4)$, where $o_{act}$ and $v_{act}$ are the numbers of active occupied and virtual orbitals, respectively. The cost of the MP2 correction for the external space is non-iterative and scales as $O(N^5)$ overall (dominated by integral transformations), with the MP2 energy evaluation itself being $O(o^2 v^3)$. Since $v_{act}$ is chosen to be a small fraction of the total virtual space (e.g., 15–25\% in our benchmarks), the overall cost is dominated by the much cheaper MP2 calculation, leading to significant computational savings--often one to two orders of magnitude--especially for large molecules and basis sets.

Overall, the active space UCCSD(4)/MP2 framework provides a practical and systematically improvable strategy for extending unitary coupled cluster methods to larger molecules and more complex chemical processes on classical computers. By explicitly separating internal and external correlation treatments and clarifying the regimes in which different orbital choices and coupling schemes are most effective, this work offers concrete guidance for applying UCC-based approximations in practice. Furthermore, by reducing the size of the active orbital space, this framework offers a viable path toward scaling unitary coupled cluster calculations for implementation on resource-constrained quantum hardware, bridging the gap between current algorithms and future chemical applications.

\begin{acknowledgement}
P.V. and B.R. were funded by NSF CTMC CAREER Award 2046744. This research was conducted using computational resources and services at the Center for Computation and Visualization, Brown University.
\end{acknowledgement}

\section*{Supporting Information}
Supplemental PDF: Includes a detailed molecule-by-molecule analysis of the UCCSD(4)/MP2 performance on the GW100 dataset, including absolute and percentage error distributions represented through violin plots and heatmaps. All custom codes used in this work are open-source and available on GitHub at \nolinkurl{https://github.com/prateekvaish/est.git}.


\bibliography{main}




\end{document}


\section{Performance on Species in Their Equilibrium Geometries}

This section extends the analysis presented in Section 4.1 of the main text by evaluating the performance of the active space UCCSD(4)/MP2 methods against the CCSD(T). While the main text primarily utilizes UCCSD(4) as an internal reference to assess the active space approximation itself, these results contextualize the absolute accuracy of the composite (c-) and interacting (i-) variants within the broader framework of high-level electronic structure theory. We report absolute errors in total energies for a representative set of small molecules--LiH, CH$_4$, H$_2$O, HF, N$_2$, F$_2$, and H$_2$CO--using both cc-pVDZ and cc-pVTZ basis sets with 60\% of the virtual orbitals retained in the active space.

\begin{figure}[ht]
    \centering
    \begin{subfigure}[t]{0.49\linewidth}
        \centering
        \includegraphics[width=\linewidth]{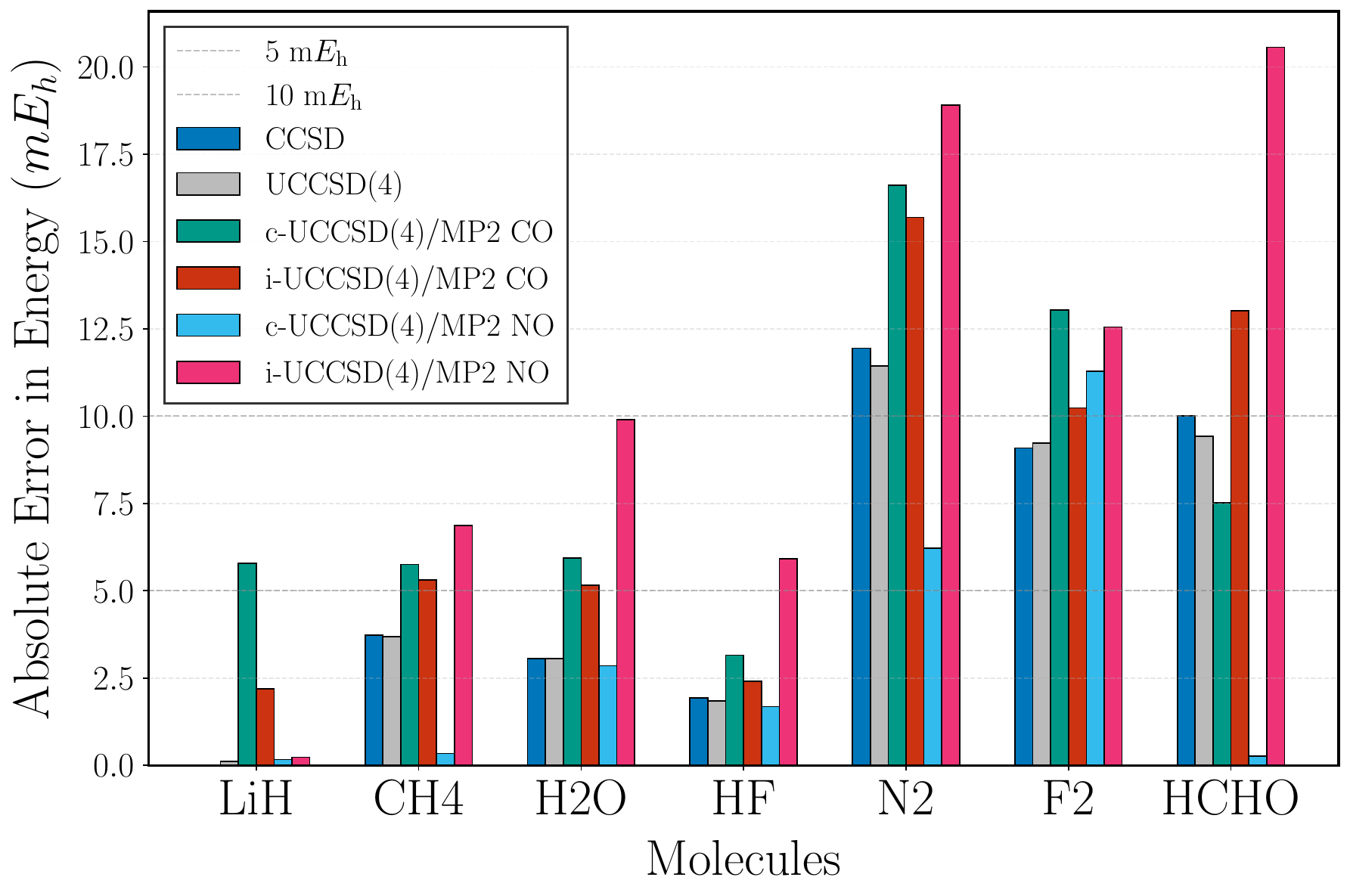}
        \caption{cc-pVDZ basis}
        \label{fig: dz equilibrium dataset}
    \end{subfigure}
    \hfill
    \begin{subfigure}[t]{0.49\linewidth}
        \centering
        \includegraphics[width=\linewidth]{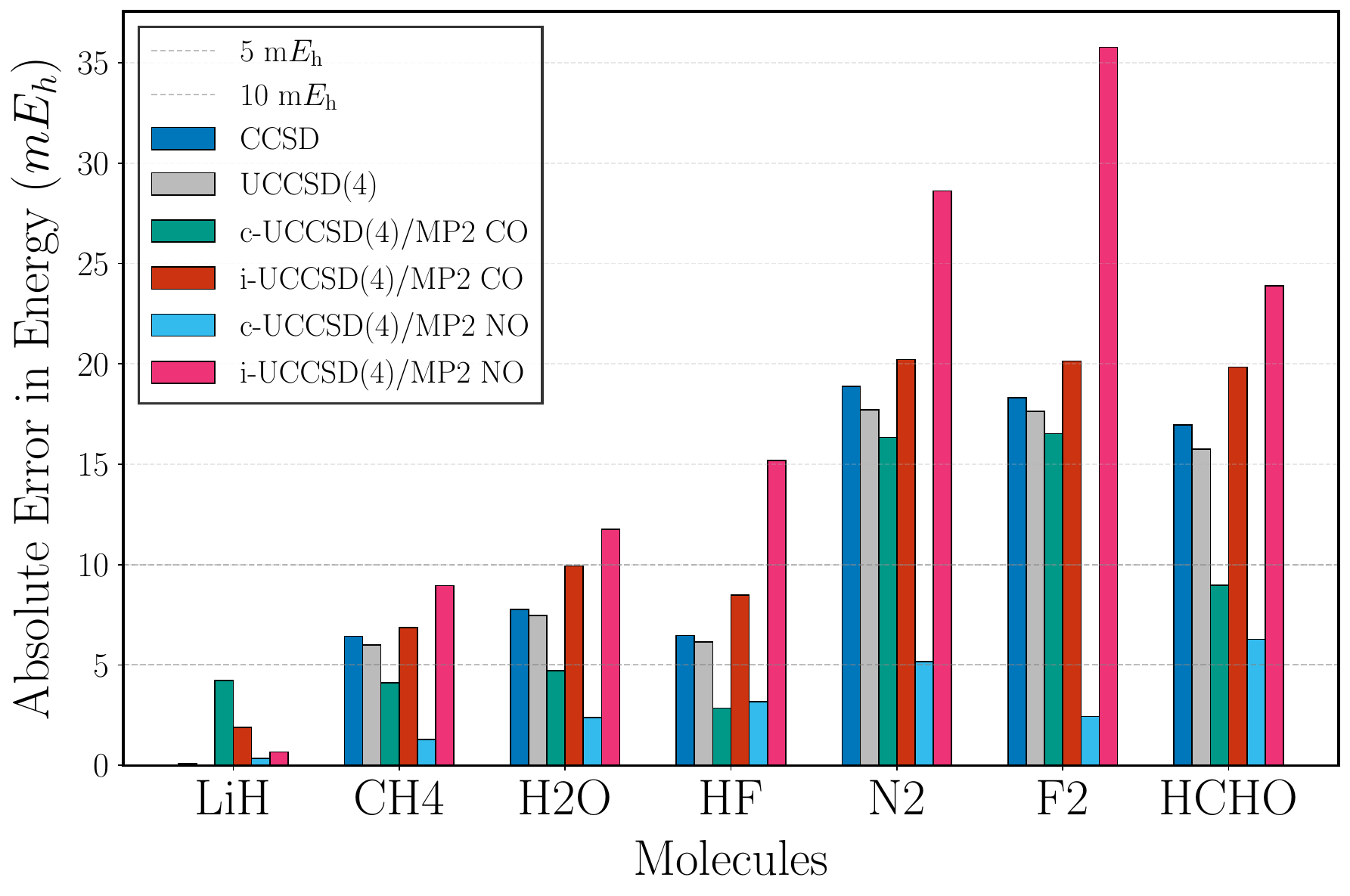}
        \caption{cc-pVTZ basis}
        \label{fig: tz equilibrium dataset}
    \end{subfigure}
    \caption{Absolute errors in total energies (in millihartrees) relative to the CCSD(T) reference for a range of molecules. Each bar represents a different approximation—CCSD, composite (c-UCCSD(4)/MP2), and interacting (i-UCCSD(4)/MP2)—in combination with canonical orbitals (CO), or natural orbitals (NO). Only 60\% of the virtual orbitals are retained in the active space.}
    \label{fig: equilibrium_dataset_combined}
\end{figure}

\section{Detailed Performance on the GW100 Dataset}

\begin{figure}[ht]
    \centering
    \begin{subfigure}[t]{0.49\linewidth}
        \centering
        \includegraphics[width=\linewidth]{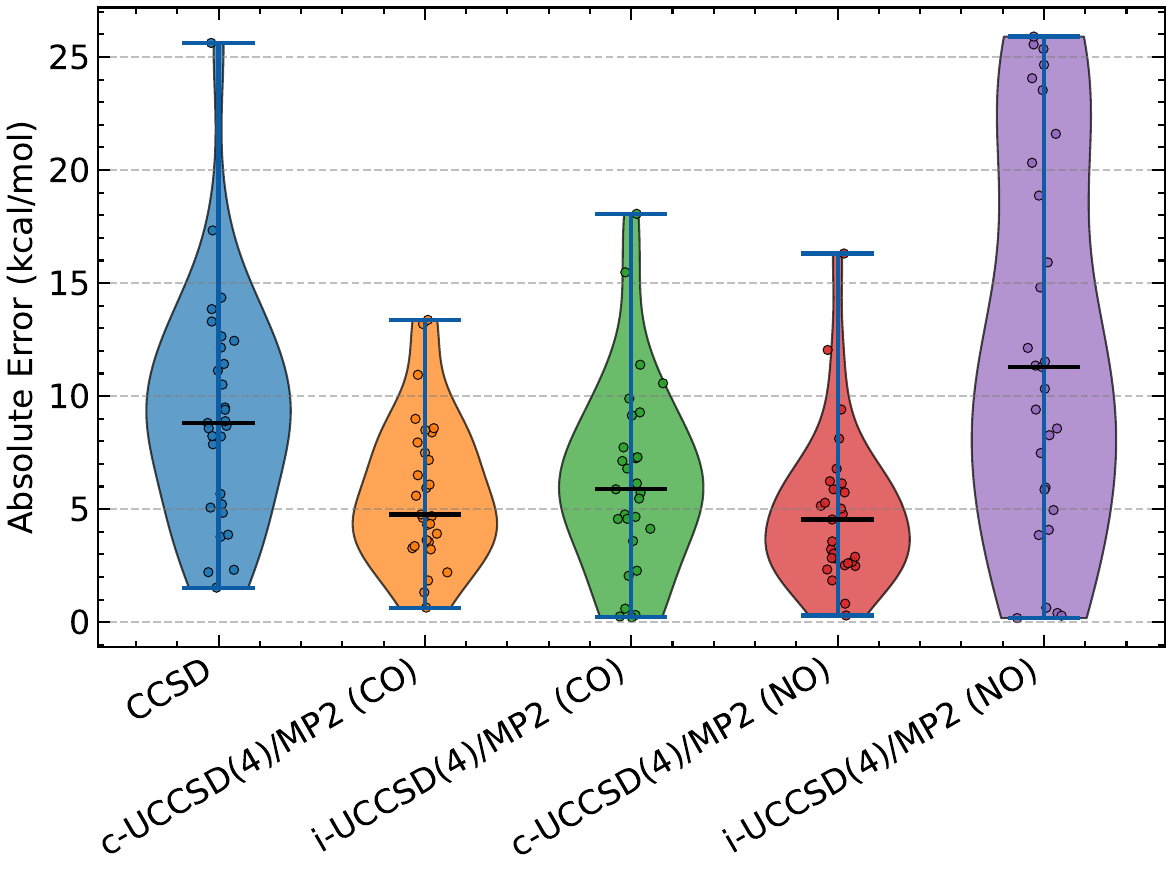}
        \caption{Medium benchmark set}
        \label{fig: medium_violin}
    \end{subfigure}
    \hfill
    \begin{subfigure}[t]{0.49\linewidth}
        \centering
        \includegraphics[width=\linewidth]{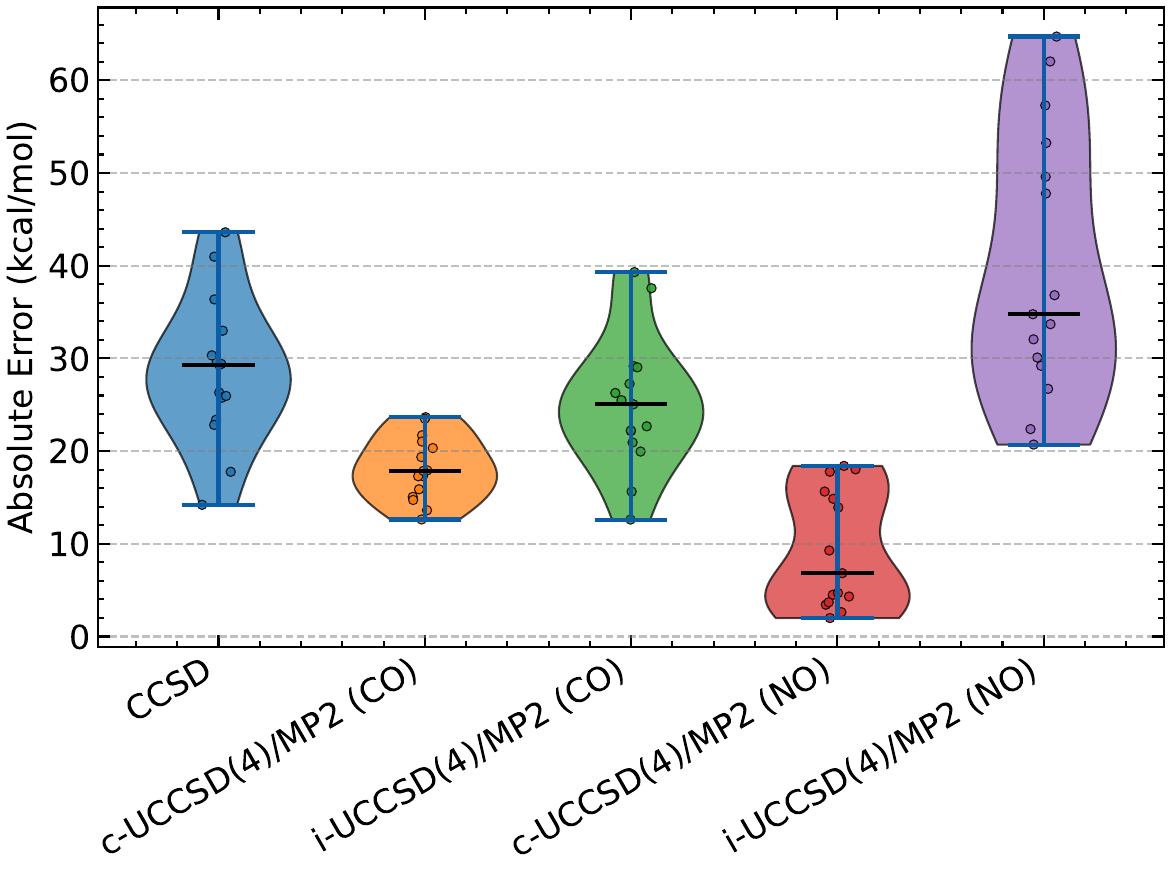}
        \caption{Large benchmark set}
        \label{fig: large_violin}
    \end{subfigure}
    \caption{Absolute error in correlation energies (in mEh) with respect to the
    CCSD(T) reference for (a) medium‐sized and (b) large molecules.  Each
    violin envelope shows the distribution of errors over all molecules in
    the set; the thick horizontal bar marks the median and the whiskers span
    the full range, while the overlaid points correspond to individual
    molecules. ''CCSD'' is the full canonical CCSD result.  All remaining
    methods employ the active–space \mbox{UCCSD(4)/MP2} method.  ''CO'' and ''NO'' indicate
    that the calculations were performed in canonical orbitals and natural
    orbitals, respectively.}
    \label{}
\end{figure}

\begin{figure*}[htbp]  
  \centering
  \begin{subfigure}[t]{0.48\textwidth}
    \includegraphics[width=\linewidth]{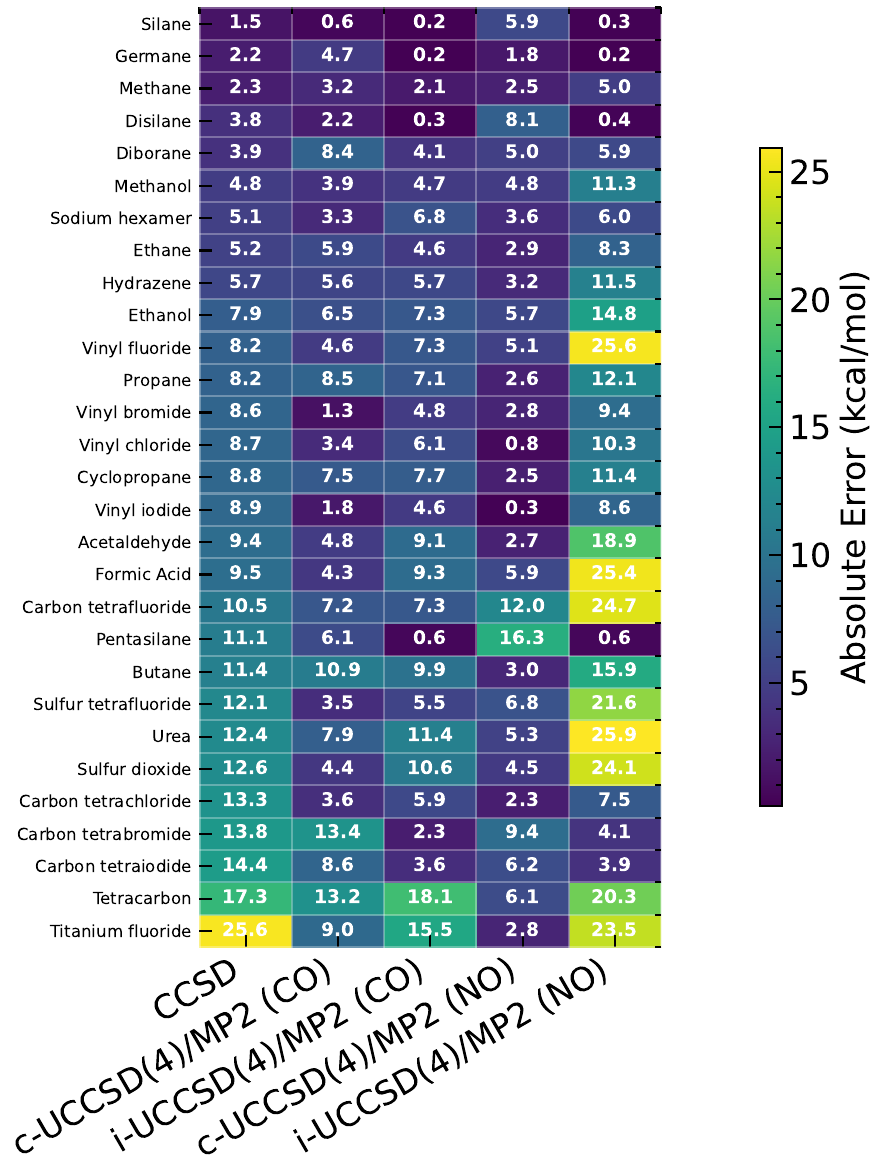}
    \caption{Medium benchmark set}
    \label{fig:med_heatmap}
  \end{subfigure}
  \hfill
  \begin{subfigure}[t]{0.48\textwidth}
    \includegraphics[width=\linewidth]{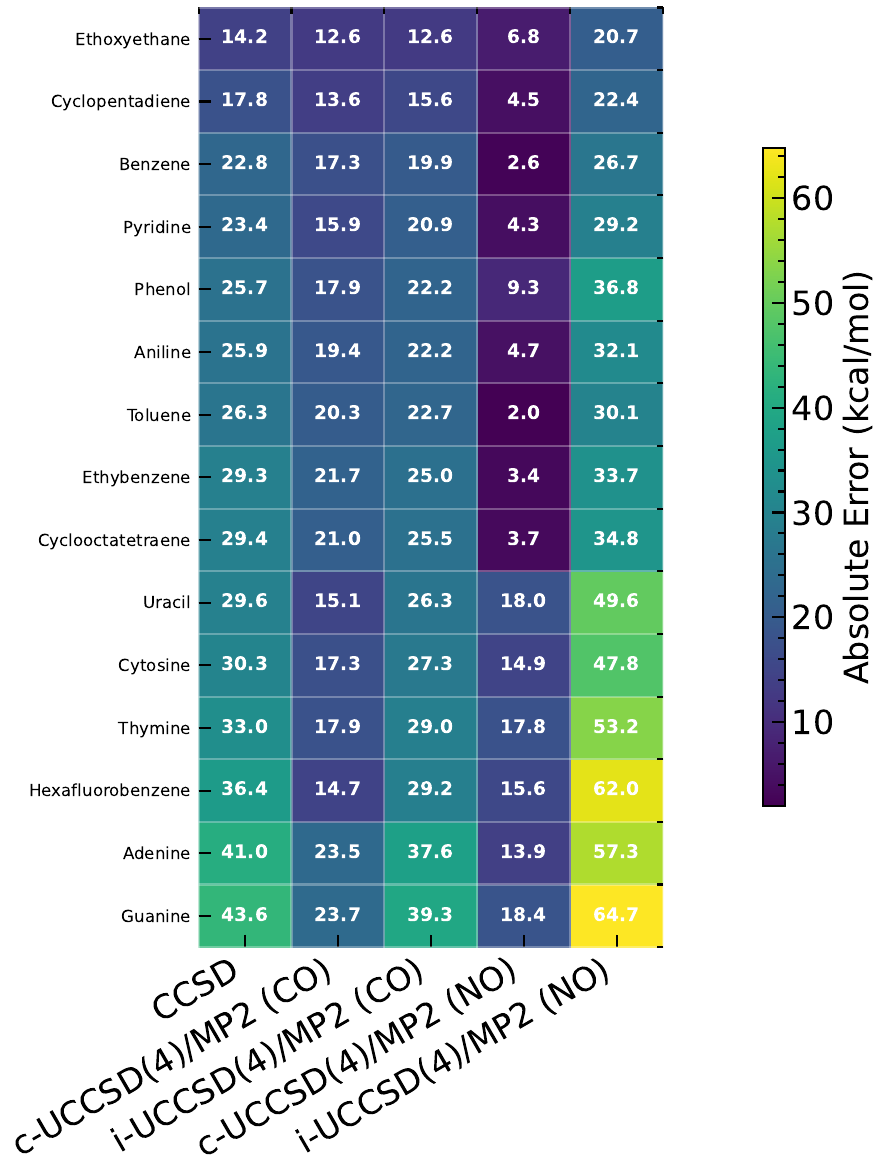}
    \caption{Large benchmark set}
    \label{fig:large_heatmap}
  \end{subfigure}

  \caption{Absolute error (in mEh) in correlation energies
    with respect to the CCSD(T) reference for (a) medium‐sized and (b) large
    molecules.  Each row corresponds to a single molecule and each column to
    a quantum-chemical method.  Numerical values are printed inside the
    cells and the colour scale to the right emphasises the magnitude of the
    error. ''CCSD'' denotes the full canonical CCSD result.  All other
    columns correspond to active–space \mbox{UCCSD(4)/MP2} methods. ''CO'' and ''NO'' indicate canonical orbitals and natural orbitals, respectively.}
  \label{fig:error_heatmaps}
\end{figure*}

\begin{figure*}[htbp]  
  \centering
  \begin{subfigure}[t]{0.48\textwidth}
    \includegraphics[width=\linewidth]{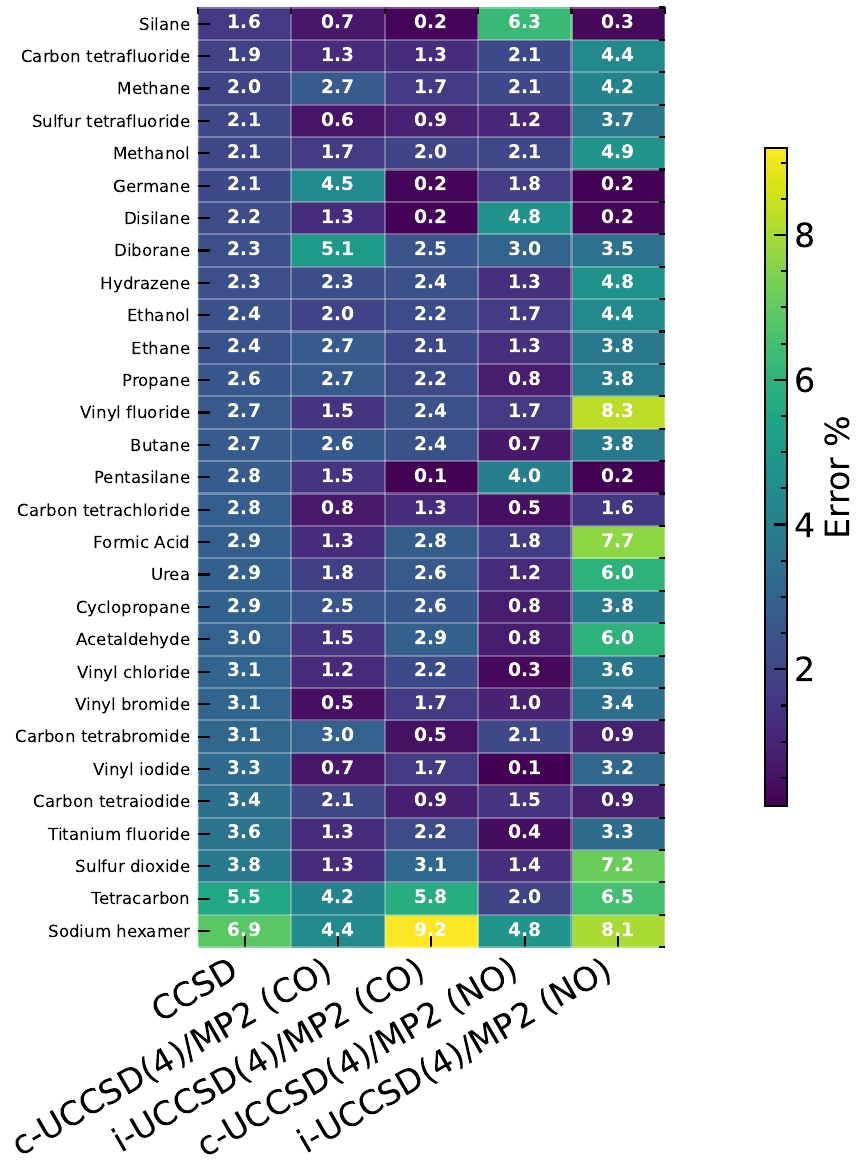}
    \caption{Medium benchmark set}
    \label{fig:med_heatmap}
  \end{subfigure}
  \hfill
  \begin{subfigure}[t]{0.48\textwidth}
    \includegraphics[width=\linewidth]{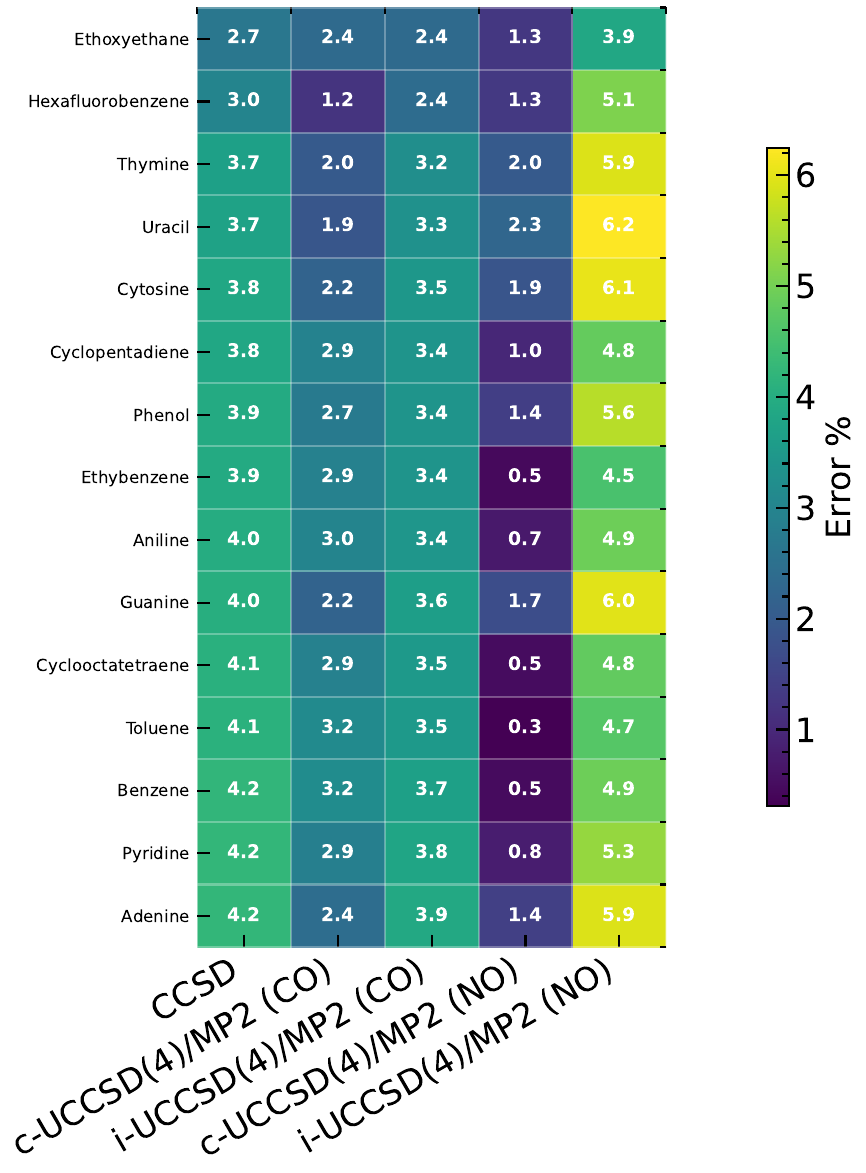}
    \caption{Large benchmark set}
    \label{fig:large_heatmap}
  \end{subfigure}

  \caption{Percentage error in correlation energies with respect to the CCSD(T) reference for (a) medium‐sized and (b) large molecules. Each row corresponds to a single molecule and each column to a quantum-chemical method.  Numerical values are printed inside the cells and the colour scale to the right emphasises the magnitude of the error. ''CCSD'' denotes the full canonical CCSD result.  All other columns correspond to active–space \mbox{UCCSD(4)/MP2} methods. ''CO'' and ''NO'' indicate canonical orbitals and natural orbitals, respectively.}
  \label{fig:error_pct_heatmaps}
\end{figure*}

This section provides a detailed, molecule-by-molecule analysis of the performance of the active space UCCSD(4)/MP2 methods on the GW100 dataset, complementing the summary presented in the main text. We report absolute errors in millihartrees (mEh) and percentage errors relative to the reference CCSD(T) correlation energies.
